\documentclass{emulateapj}

\newcommand{\CL}{RX J0152.7-1357}
\newcommand{\myemail}{rdemarco@astro-udec.cl}

\shorttitle{Star formation histories in a cluster environment at $z\sim0.84$}
\shortauthors{Demarco et al.}

\begin{document}

\title{Star Formation Histories in a Cluster Environment at $z\sim0.84$}

\author{R. Demarco\altaffilmark{1,2}, R. Gobat\altaffilmark{3},
  P. Rosati\altaffilmark{4}, C. Lidman\altaffilmark{5},
  A. Rettura\altaffilmark{6}, M. Nonino\altaffilmark{7}, A. van der
  Wel\altaffilmark{8}, M. J. Jee\altaffilmark{9},
  J.~P.~Blakeslee\altaffilmark{10}, H. C. Ford\altaffilmark{2},
  M. Postman\altaffilmark{11}} \affil{}

\altaffiltext{1}{Department of Astronomy, Universidad de Concepci\'on. Casilla 160-C, Concepci\'on, Chile\\ \myemail}
\altaffiltext{2}{Department of Physics \& Astronomy, The Johns Hopkins University, 3400 N. Charles Street. Baltimore, MD 21218, USA}
\altaffiltext{3}{CEA, Laboratoire AIM, Irfu/SAp, F-91191 Gif-sur-Yvette, France}
\altaffiltext{4}{European Southern Observatory, Karl-Schwarzschild-Strasse 2, D-85748 Garching, Germany}
\altaffiltext{5}{Anglo Australian Observatory, P.O. Box 296, Epping NSW 1710, Australia}
\altaffiltext{6}{Department of Physics \& Astronomy, University of California, Riverside. 900 University Ave. Riverside, CA 92521, USA}
\altaffiltext{7}{INAF-OAT, via G.B. Tiepolo 11, 40131 Trieste, Italy}
\altaffiltext{8}{Max-Planck Institute for Astronomy, K\"onigstuhl 17, D-69117, Heidelberg, Germany}
\altaffiltext{9}{Department of Physics, University of California, Davis, One Shields Avenue, Davis, CA 95616, USA}
\altaffiltext{10}{Herzberg Institute of Astrophysics, National Research Council of Canada, Victoria, B.C. V9E 2E7, Canada}
\altaffiltext{11}{Space Telescope Science Institute, Baltimore, MD 21218, USA}

\begin{abstract}

We present a spectrophotometric analysis of galaxies belonging to the
dynamically young, massive cluster \CL\ at $z\sim0.84$, aimed at
understanding the effects of the cluster environment on the star
formation history (SFH) of cluster galaxies and the assembly of the
red-sequence (RS). We use VLT/FORS spectroscopy, ACS/WFC optical and
NTT/SofI near-IR data to characterize SFHs as a function of color,
luminosity, morphology, stellar mass, and local environment from a
sample of 134 spectroscopic members. In order to increase the
signal-to-noise, individual galaxy spectra are stacked according to
these properties. Moreover, the D4000, Balmer, CN3883, Fe4383 and
C4668 indices are also quantified. The SFH analysis shows that
galaxies in the blue faint-end of the RS have on average younger stars
($\Delta t \sim 2$ Gyr) than those in the red bright-end. We also
found, for a given luminosity range, differences in age ($\Delta t
\sim 0.5 - 1.3$ Gyr) as a function of color, indicating that the
intrinsic scatter of the RS may be due to age variations. Passive
galaxies in the blue faint-end of the RS are preferentially located in
the low density areas of the cluster, likely being objects entering
the RS from the ``blue cloud''. It is likely that the quenching of the
star formation of these RS galaxies is due to interaction with the
intracluster medium. Furthermore, the SFH of galaxies in the RS as a
function of stellar mass reveals signatures of ``downsizing'' in the
overall cluster.

\end{abstract}


\section{Introduction}

There is no doubt that galaxy evolution is influenced by the local
environment. The observed properties of galaxies (color, magnitude,
morphology, metallicity) are associated with their local
neighborhood. The latter is usually characterized in terms of the
local number density of galaxies.

One of the most prominent connections between galaxy properties and
environment is the morphology-density relation \citep{d80}, by which
early-type galaxies dominate high-density environments in contrast to
late-type galaxies that dominate low-density ones. This relation has
been quantified up to $z\sim1$, showing different evolutionary
patterns depending on whether the galaxy sample is selected based on
luminosity \citep{pfc05,ste05} or stellar mass
\citep{hif07,vhf07}. 

For mass selected samples, \citet{vhf07} show that galaxies have to
evolve in mass, morphology, and density such that the
morphology-density relation does not change since at least
$z\sim0.8$. In the case of luminosity-selected samples, the
morphology-density relation observed at $z\sim1$ \citep{pfc05} is
reported to hold up to $z\sim1.46$ \citep{hss09}.

Since early-type galaxies are among the reddest objects in any given
sample at a given epoch and contain the bulk of the stellar mass in
the Universe, the above morphology-density relation can be translated
into two other relations: color-density and stellar mass-density. In
particular, the color-density relation, characterized by the tendency
of red galaxies (mostly early-type ones) to be found in the core of
clusters, can be used as a tool to identify high-redshift clusters. It
takes advantage of one of the most distinctive features in the
color-magnitude diagram of a cluster: the so-called red-sequence
\citep*[RS;][]{d61,vs77}.

This RS, however, is not exclusive of clusters as it is also found in
the field. In fact, some of the observed properties of the RS such as
its color scatter and luminosity coverage vary, at a given redshift,
with local galaxy density \citep*[e.g.,][]{tka05}. This highlights the
influence that the local environment has on galaxy properties such as
colors.

It has been observed that the RS of clusters has a small scatter in
color and a slope which do not seem to evolve over $\sim9$ Gyr of
cosmic time since $z\sim1.4$
\citep*[e.g.,][]{sed98,bfp03,mhb06b,lrt08}. This lack of evolution was
shown to be better explained by a RS being primarily a
color-metallicity relation instead of a color-age one
\citep{ka97}. However, more recent evidence gathered from local
samples of galaxies shows that variations of stellar age along the
red-sequence are also present \citep*[e.g.,][]{gcb06,bns06} in
addition to variations in metallicity, with some of the intrinsic
color scatter of the red-sequence being due to stellar age differences
\citep*[see also][]{tfi07}.

A common procedure is to use the spectral energy distribution (SED)
fitting technique to estimate ages and formation redshifts for cluster
galaxies in the RS
\citep*[e.g.,][]{bfp03,lrd04,lrt08,bhf06,mbs06a,mhb06b,grs08,rrn08}. This
procedure allow us to constrain the epoch and mode of formation of
massive early-type galaxies. When spectroscopic data allow it,
spectral indices are also measured to constrain the properties and
evolution of cluster galaxies \citep*[e.g.,][]{jbd05,tfi07,bpb09}.

At $z=1.2$, \citet{grs08} find that the bulk of the stars in cluster
early-type galaxies is formed $\sim$ 0.5 Gyr earlier than that in
field early-type galaxies, and RS galaxies were already in place
$\sim1$ Gyr earlier. Although the most massive (in stellar content)
early-type galaxies do not show such an age difference with
environment, this age divergence is most noticeable at stellar masses
$\lesssim10^{11} M_{\odot}$. A similar conclusion was reached by
\citet{rrn08}, who show that the environment regulates the timescale
associated with the SFHs of early-type galaxies, with a fraction of
field system having a more extended period of stellar mass assembly.

\citet{bpb09} find that the SFHs of galaxies in two clusters at
$z\sim0.3$ depend on local environment which is also related to the
cluster dynamical state. In addition to the expected gradient of star
formation with clustercentric distance, both luminous ($L\geq L^*$)
and sub-luminous members contribute to a sharp increase of the star
formation activity along filaments connected to the dynamically young,
merging system. The more relaxed cluster, on the other hand, is mostly
dominated by red, passive galaxies or galaxies whose star formation is
being quenched.

The novel procedure used by \citet{grs08} to determine the SFH of
cluster galaxies combines both the SED fitting technique and,
simultaneously, a fit to the spectroscopic features (pseudo-continuum
and absorption) of a galaxy spectrum. The SED covers a wide range of
wavelengths and provides information on mass and current SFH while the
spectrum, although on a much more limited wavelength range, allows one
to determine the age of the stellar population of a galaxy with
greater precision. The combination of photometry and spectroscopy
therefore puts stronger constraints on the SFH than either alone. This
approach can also be complemented by determining some relevant
spectral indices available at the observed wavelength range.

The galaxy cluster \CL\ \citep{dsg00} at $z\sim0.84$, with its
dynamically young and complex structure, represents and ideal
laboratory to study the relation between galaxy SFH and
environment. Here we apply the above spectrophotometric fitting
technique to the spectroscopically confirmed population of cluster
members. Our goal is to deepen our understanding of how galaxy
evolution is driven by environmental processes and, in particular, to
better constrain and understand the physical mechanisms that
contribute to form the RS and set its observed properties.

\CL\ is one of the most distant X-ray luminous clusters discovered in
the ROSAT Deep Cluster Survey \citep{rdn98}. It is observed at an
epoch of greater cosmic activity in terms of stellar mass buildup in
galaxies \citep*[e.g.,][]{dbs06}, with the cluster itself being
assembled by the merging of two subclusters \citep{drl05,gdr05} and by
the accretion of groups from surrounding filaments \citep{tka06}.

To date, \CL\ has been the subject of a number of studies: ICM
structure and X-ray properties \citep{mje03}, RS properties
\citep{bhf06,pkh08}, cluster dynamics and substructure
\citep{drl05,gdr05}, star-forming members \citep{hdr05},
Sunyaev-Zel'dovich properties \citep{jlg01}, weak lensing mass
structure \citep{jwb05}, physical properties of galaxy members
\citep{jbd05}, infrared sources in the cluster \citep{mrr07}, and
large-scale filaments associated with it \citep{tka06}. It is one of the
clusters in the ACS intermediate-redshift cluster program
\citep{fpb04} with one of the most comprehensive datasets, from X-ray
to infrared.

We improve over previous studies of this cluster by considering,
simultaneously, a large enough ($\sim130$) number of spectroscopic
members with high enough ($<15$ \AA) spectral resolution data and
5-band photometry (from optical to near-infrared). In addition, we
characterize the local environment by the relative, projected dark
matter density instead of the local projected number density of
galaxies. We estimate SFHs and spectral indices as a function of
color-magnitude, morphology, stellar mass and location within the
cluster, which allows us to establish a comprehensive view of galaxy
properties and their connection with the environment. In contrast to
other works at this redshift, given the available number of members in
the RS, we divide this one into a grid in color-magnitude space where
the average SFH of galaxies can be studied at different colors and
magnitudes. This approach is aimed at providing a deeper insight into
RS variations in order to unveil the physics responsible of its
observed properties such as its intrinsic scatter.

The present work is structured in the following way. In
\S\ref{observations} we describe the observations and the dataset
used in our analyses. In \S\ref{analysis} we focus on the technical
details of this investigation, such as the definition of regions in
the hyperspace of galaxy properties considered for this investigation,
the available spectra, the spectrophotometric fitting procedure used
to characterize the average SFH within those regions, and the spectral
indices used to complement such SFH characterization. In
\S\ref{results} we present the results of our analyses followed, in
\S\ref{discussion} by a discussion focused on the formation of the RS
in the cluster. We finally summarize our main conclusions in
\S\ref{conclusion}.

Throughout the paper, unless explicitly indicated, we assume a
$\Lambda$CDM cosmology with H$_0=70$ km s$^{-1}$ Mpc$^{-1}$,
$\Omega_M=0.3$ and $\Omega_{\Lambda}=0.7$.

\section{Observations}\label{observations}

\CL\ has been observed with a number of instruments from the ground
and space. In this work we have used optical and near-IR imaging data
obtained with HST/ACS and NTT/SofI, respectively, and optical
spectroscopy obtained with VLT/FORS. Descriptions of these
observations and data reductions can be found in the existing
literature \citep{drl05,bhf06}, hence, we only provide a short summary
of them below.

\subsection{Imaging and photometry}\label{imaging}

As reported in \citet{bhf06}, \CL\ was observed with ACS \citep{F98}
on HST in three bands: F625W, F775W and F850LP, hereafter referred to
as $r_{625}$, $i_{775}$ and $z_{850}$, respectively. The cluster was
imaged using a 2 $\times$ 2 overlapping pattern, producing a mosaic of
about 5\farcm8 on a side with a central overlapping region of about
1\arcmin. Each pointing was observed for two orbits per filter, giving
a total orbit expenditure of 24 orbits. The images were processed and
the final mosaic produced using the ACS GTO Apsis pipeline
\citep{bam03}. For more details, see \citet{bhf06}.

The near-IR data were obtained with SofI \citep{mcl98} on the ESO NTT
\citep*[see][]{drl05}. The cluster was imaged in the J and K$_s$ bands
under sub-arcsecond seeing for 3.8 and 3 hours, respectively. The
final images cover a region of 4\farcm9 on a side and were reduced in
a standard manner.

In order to use the ACS and SofI data in a consistent way, we produced
a new multi-band photometric catalog, different from those used in
\citet{drl05} (SofI) and \citet{bhf06} (ACS). Photometry from the ACS
data was obtained in dual mode using the ACS $z_{850}$ image for
detections. By running SExtractor \citep{ba96} on the ACS mosaic, 41
point sources were selected to derive aperture corrections. Magnitudes
\citep*[in the AB system;][]{o74} within radii of 0\farcs75 and
2\farcs0 were compared, resulting in differences of 0.039 for both the
$r_{625}$ and $i_{775}$ bands and 0.046 for the $z_{850}$ filter. Zero
points (in AB magnitudes) for the $r_{625}$-, $i_{775}$- and
$z_{850}$-band data are 36.542, 36.321 and 35.520,
respectively. 

The near-IR J and K$_s$ images were registered onto the ACS ones, with
residuals of less than 1 pixel for both J and K$_s$. The same point
sources were used to derive corrections between apertures of 1\arcsec\
and 5\arcsec\ in radius. These corrections are 0.254 and 0.239 for J
and K$_s$, respectively. Additionally, extinction corrections
\citep{sfd98} of 0.014 in J and 0.009 in K were applied, resulting in
a total correction for the 1\arcsec\ radius aperture of 0.268 in J and
0.248 in K$_s$. Zero points (in the AB system) in J and K$_s$ are
27.260 and 26.745, respectively. Hereafter, unless otherwise
indicated, magnitudes are in the AB system. To transform the near-IR
photometry from the Vega system to AB magnitudes, we used corrections
of 0.960 and 1.895 for the J and K$_s$ bands.

For completeness, we note that flanking fields surrounding the central
$r_{625}$, $i_{775}$ and $z_{850}$ mosaic were obtained with ACS in
F606W (broad V) and F814W (broad I), however, we did no attempt to use
those data in this work because of their shallower integration,
reduced photometric coverage, and lack of overlap with the existing
near-IR data.

\subsection{Spectroscopy}

\CL\ and its outskirts have been subject of a number of extensive
spectroscopic surveys \citep{drl05,jbd05,tka06,pkh08}. While the
survey by \citet{tka06} concentrated on the large-scale structures
surrounding the main cluster, only 8 members in J{\o}rgensen's survey
were not included in Demarco's work. In the present study, we use the
spectra of the 102 cluster members confirmed by \citet{drl05},
complemented with the spectra of 32 new cluster members obtained from
subsequent FORS2 \citep{ar92} spectroscopy on the ESO VLT. These 134
sources are listed in table \ref{tab_membs}. IDs are in the same
system of those in \citet{drl05}, and the last column corresponds to
their emission line flag: a value of 0 is given to passive galaxies, a
value of 1 is given to emission line galaxies, and a value of 2 is
given to AGN.

The new 32 spectroscopic members were obtained after targeting
clusters candidates with 4 multi-object masks, using the Mask Exchange
Unit (MXU) on FORS2, and exposing each of them until reaching
integration times between 3 and 4 hours. The data were collected
between November 4th and November 7th, 2005, in Visitor Mode (ESO
program ID 076.A-0889(A)) under an average seeing of $\sim0\farcs9$. A
total of 152 galaxies were observed with slits of 1\arcsec\ in width
using the 300V grism. The data were binned by 2 pixels which resulted
in a dispersion of $\sim 3.3$\AA/pixel and a spectral resolution of
$\sim 13$\AA. One mask was observed using no order-separation filter
while the other three masks were exposed using the GG375+80
filter. Although these observations were designed to target
gravitational arc candidates, the new members resulted from putting
the slits on fillers that were likely cluster members based on their
photometric redshift \citep*[$0.7 < z_{phot} < 0.95$; see][]{drl05}.

The observations were prepared and the data reduced in a similar way
and using the same dedicated software as described in
\citet{drl05}. Redshifts were obtained by cross-correlating
\citep{td79,kmw92} the observed spectra with template galaxy spectra
from \citet{kcb96}. Out of 113 redshifts, 32 were securely confirmed
within the range defining cluster membership \citep*[$0.81 < z <
  0.87$; see][]{drl05}. Observational errors (obtained from observing
several sources more than once) are of the same order of magnitude as
those reported in \citet{drl05,drl07}, i.e., $\delta z\sim 8-12\times
10^{-4} \sim 10^{-3}$.

\section{Analysis}\label{analysis}

In an effort to better understand how galaxy properties, such as age
and stellar mass, contribute to establishing the observed
characteristics of the RS, we focus the present analysis on the star
formation history (SFH) of cluster galaxies in the RS. As the
luminosity coverage and scatter of the RS are observed to vary between
cluster and field environment \citep*[e.g.,][]{tka05}, we also
investigate the impact that the intracluster environment may have on
the above SFHs. 

Traditionally, SFHs are determined by fitting synthetic galaxy spectra
to the available photometry \citep*[e.g.,][]{rrs06,rrn08}, however, in
the present analysis, we additionally perform a simultaneous fit to the
available spectra as performed by \citet{grs08}. In order to increase
signal-to-noise (S/N) for the spectrophotometric fitting, galaxy
spectra are co-added as explained in \S\ref{stacking}. Before
stacking, galaxy spectra are grouped according to color, magnitude,
location within a given subcluster and with respect to the projected
DM distribution, stellar mass, and visual morphology. All these
grouping regions are defined in \S\ref{indx_regs} and the details of
the spectrophotometric fitting procedure are given in \S\ref{sed_fit}.

\subsection{Co-added spectra}\label{stacking}

The individual spectra of the 134 cluster members of \CL\ used in the
present analysis have S/N ratios in the range 1 to 33 with a mean
value of 7.6. The S/N ratios were obtained from the ratio between the
mean flux and the r.m.s. flux calculated within the wavelength
intervals defining the continuum windows for the H$\delta_A$ feature
\citep{wo97}. Since the redshift survey presented in \citet{drl05} was
designed to mainly provide redshifts, the quality of the individual
spectra is not good enough to perform a meaningful fit to the
different spectral features sensitive to the SFH. Therefore, in order
to increase the S/N ratio to obtain a satisfactory SFH
characterization, we decided to use average spectra obtained from
co-adding individual spectra that were grouped according to various
criteria (see \S\ref{indx_regs}).

This stacking technique has already been successfully used in previous
studies about the stellar populations in cluster galaxies at
intermediate-redshift \citep*[e.g.,][]{dop04}, and the algorithm
employed here is the same as in \citet{drl07} and \citet{grs08}. The
individual spectra can be weighted by their S/N, and only those with a
S/N $>3$ were selected for stacking. The S/N ratios of the final,
co-added spectra vary between $\sim$6 and $\sim$42 for the unweighted
stacking, and between $\sim$6 and $\sim$62 for the weighted stacking.

\subsection{Grouping cluster members}\label{indx_regs}

SFHs are determined from co-added spectra grouped according to
relevant observables. In this work we study their dependence on galaxy
color and magnitude, projected angular distribution, visual
morphology, stellar mass, and projected DM density. These observables
can be considered as forming an hyperspace in which galaxies are
located. The regions within this hyperspace used for stacking spectra
are defined in what follows and table \ref{tab_regs} summarizes all
these definitions.

\subsubsection{Galaxy colors and luminosities}\label{col_lum_regs}

\begin{figure}
\plotone{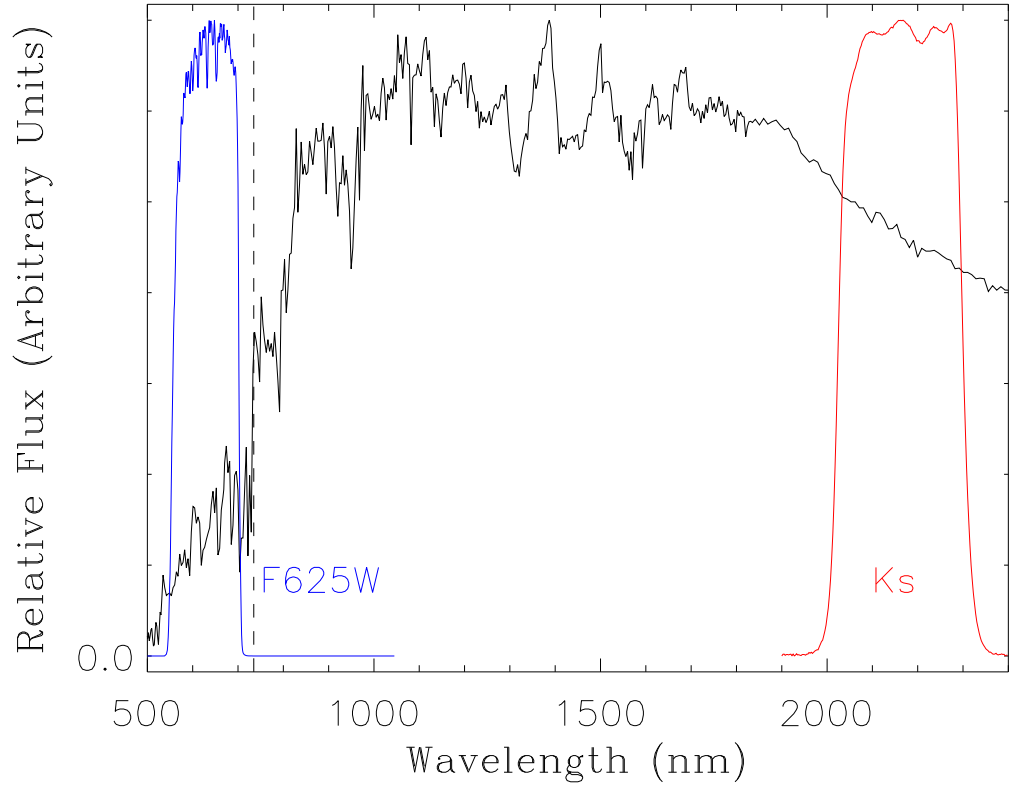}
\caption{Band-pass filters used in this work to select red-sequence
  galaxies. As a reference, a simple stellar population, 12 Gyr old,
  solar metallicity spectral energy distribution from the
  Bruzual-Charlot \citep{bc03} library, redshifted to the cluster
  redshift ($z=0.837$), is shown. The chosen filters straddle the
  rest-frame 4000\AA-break (vertical, dashed line) and the $K_s$ band
  provides coverage of the spectrophotometric region dominated by the
  bulk of the stellar content in early-type
  galaxies.\label{spec_filters}}
\end{figure}

Since we are interested in understanding the physical origin of the
scatter of the RS in clusters, and ultimately the details of the
color-luminosity evolution of cluster galaxies into the RS, we need
first to separate cluster members into blue and red galaxies. We
follow the traditional way of using two filters that straddle the
4000\AA-break at the cluster redshift to achieve
this. Fig. \ref{spec_filters} shows a simple stellar population (SSP),
12 Gyr old, solar metallicity spectral energy distribution from the
Bruzual-Charlot \citep{bc03} library, redshifted to the cluster
redshift \citep*[$z=0.837$;][]{drl05}. On top of it, our choice of
filters ($r_{625}$ and K$_s$) to straddle the 4000\AA-break is laid
out, which allows us to separate in an efficient way red (early-type)
from blue (late-type) galaxies. In contrast to \citet{bhf06}, we
prefer to use the K$_s$-band as a the ``red'' filter because of its
ability to trace the rest-frame near-IR light coming from the bulk of
the stellar content of galaxies at $z\sim0.8$, unaffected by biases
due to recent star formation \citep{sed98}.

\begin{figure}
\plotone{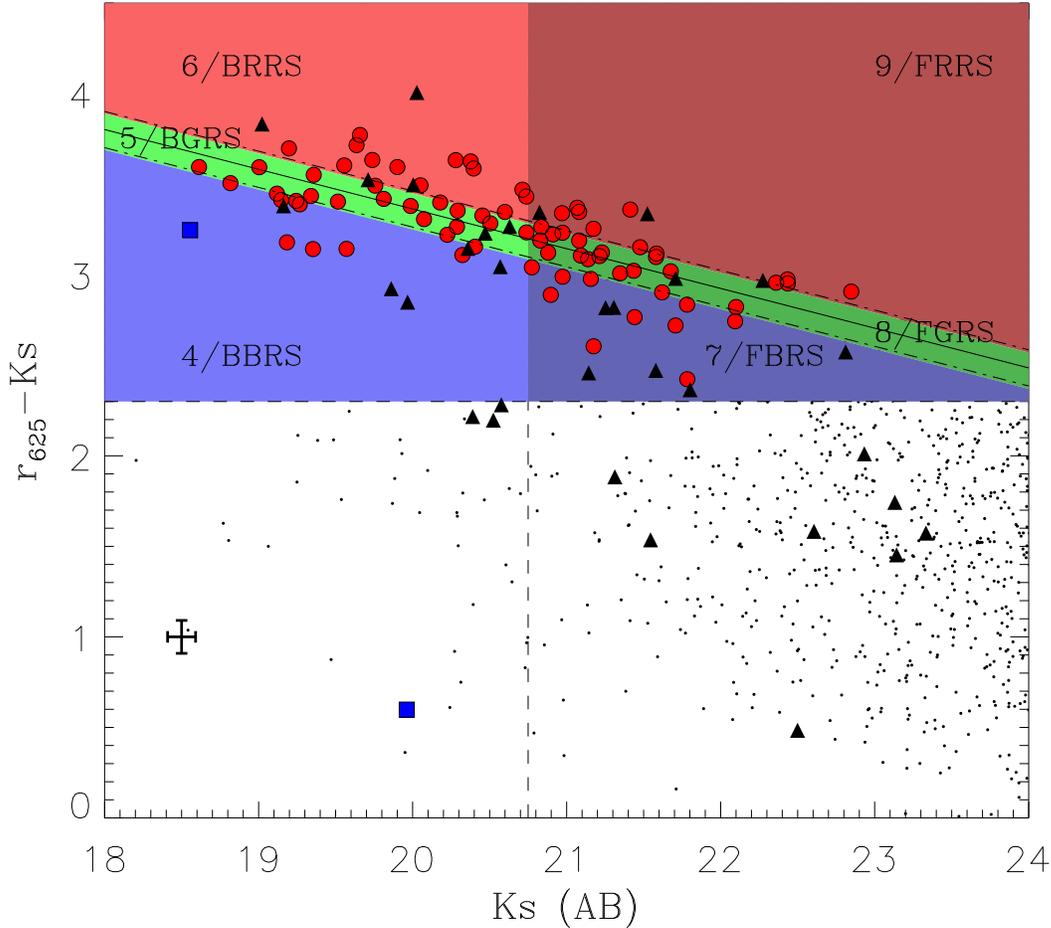}
\caption{Definition of regions in color-magnitude space used to stack
  spectra within the red-sequence (RS). The horizontal dashed line has
  arbitrarily been set at $r_{625}-Ks=2.3$ to separate blue from red
  galaxies. We consider galaxies with a $r_{625}-Ks > 2.3$ color as
  galaxies within the cluster RS. The vertical dashed line has
  arbitrarily been set to $Ks=20.75$ ($\sim K^*+1$) to divide the RS
  into ``bright'' and ``faint'' bins. The RS is thus divided into 3
  bright and 3 faint regions. The three bright regions, named 4/BBRS,
  5/BGRS and 6/BRRS, are defined in table \ref{tab_regs}. The three
  faint regions, 7/FBRS, 8/FGRS and 9/FRRS, are also defined in table
  \ref{tab_regs}. The fit to the red sequence is indicated by the
  solid black line, while the dot-dashed black lines are at $\pm0.1$
  in ($r_{625}-Ks$) from the fit. The cross at the lower-left of the
  plot indicates typical error bars in magnitude and color. Red
  circles correspond to passive cluster members, black triangles
  correspond to star-forming members, and the two blue squares are the
  confirmed AGN members. The black dots are sources within the ACS
  mosaic for which photometric information is available. Star
  formation histories within these regions allows us to study
  variations of the stellar content of cluster galaxies as a function
  of galaxy color and luminosity.\label{col_mag_regs}}
\end{figure}

Fig. \ref{col_mag_regs} shows the Color-Magnitude Diagram (CMD) of
\CL\ \citep*[for a detailed analysis of the CMD of this cluster,
  see][]{bhf06}. Red circles correspond to passive (no detectable
emission features) cluster members, black triangles correspond to
star-forming (with detectable [O$\mathrm{II}$]) members, and the two
blue squares are the confirmed AGN members \citep{drl05}. The black
dots are sources within the ACS mosaic for which photometric
information is available, and the cross at the lower-left of the plot
indicates typical error bars in magnitude and color. The horizontal
dashed line has arbitrarily been set at $r_{625}-K_s=2.3$ to separate
blue from red galaxies. We consider galaxies with a $r_{625}-K_s >
2.3$ color as belonging to the cluster RS.

In order to better explore SFH variations as a function of color and
magnitude within the RS, we have subdivided the latter into a number
of regions\footnote{Throughout the text we use the pair N/Acronym to
  identify the different stacking regions. N corresponds to the region
  ID listed in the first column of table \ref{tab_regs} and the
  Acronym is given in the comment column of the same table. The goal
  of this is to make figures easier to read while giving a physical
  meaning to the IDs at the same time.}. A first subdivision is
defined, consisting of three bins in K$_s$ magnitude, 1/BRS, 2/MRS and
3/FRS, as defined in table \ref{tab_regs}. In addition, a second and
finer subdivision, both in color and magnitude, is established, as
shown in Fig. \ref{col_mag_regs}.  The vertical dashed line has
arbitrarily been set at $K_s=20.75$ in order to divide the RS into
``bright'' and ``faint'' bins. For comparison, K$^*\sim 19.7$ (AB) at
$z\sim0.84$ \citep*[see][]{ej04}, therefore, this separation
corresponds to $\sim K^*+1$.

The locus of the RS in Fig. \ref{col_mag_regs} is obtained by a linear
least-squares fit to the data in color-magnitude space with
$r_{625}-K_s>2.3$ and $18<K_s<24$. This fit to the RS
($r_{625}-K_s=(-0.219\pm0.001)\times K_s+(7.751\pm0.026)$), indicated
by the solid black line, is used to define the ``blue'', ``green''
and ``red'' areas parallel to it in color-magnitude space, further
separated into 3 bright (regions 4/BBRS, 5/BGRS and 6/BRRS,
respectively) and 3 faint (regions 7/FBRS, 8/FGRS and 9/FRRS,
respectively) bins (see table \ref{tab_regs} for their definitions).

Only passive galaxies (red circles) within these 9 regions were
considered for stacking, because of our interest in focusing on the
quiescent, early-type galaxy population. In general red, star-forming
sources are dust-enshrouded systems
\citep*[e.g.,][]{smg99,wgm05}. However, we note that a few of the
apparently passive, RS galaxies could in fact be star-forming systems
with the [O$\mathrm{II}$] feature totally suppressed by a large amount
of dust \citep*[e.g.,][]{smg99}.

The above color separation for galaxies in the RS does not follow the
slope of the fit to the RS. Selecting galaxies following this slope
would tend to include more galaxies into the blue faint-end of the RS
that may belong to the so-called ``green valley'' or to the ``blue
cloud''. Since we are interested in passive galaxies in the RS, the
flat color separation we have adopted produces no different result
from a slope-driven color separation.

We have 76 spectra corresponding to non [O$\mathrm{II}$] emission line
galaxies in the RS available for our analyses. In order to have a
reasonable number of sources to be co-added, we set the width of the
two (bright and faint) central regions (green hatched areas) of the
finer partition to be 0.2 mag in $r_{625}-Ks$ ($\pm0.1$ in
$r_{625}-Ks$ from the fit), which is about $1.1$ times the observed
r.m.s. color scatter.

\subsubsection{Projected angular distribution}\label{angdist}

To investigate the relation between stellar content and local
environment, we study the SFH of cluster galaxies as a function of
angular distribution on the sky. Due to the complex matter (DM, gas
and galaxies) distribution of \CL\ \citep{mje03,drl05,jwb05,gdr05}, it
is very difficult to determine the center of the cluster. Instead of
stacking galaxies in concentric rings with a common center, we try the
following approach that considers the known main subclusters
\citep*[see][]{drl05,gdr05}.

\begin{figure}
\plotone{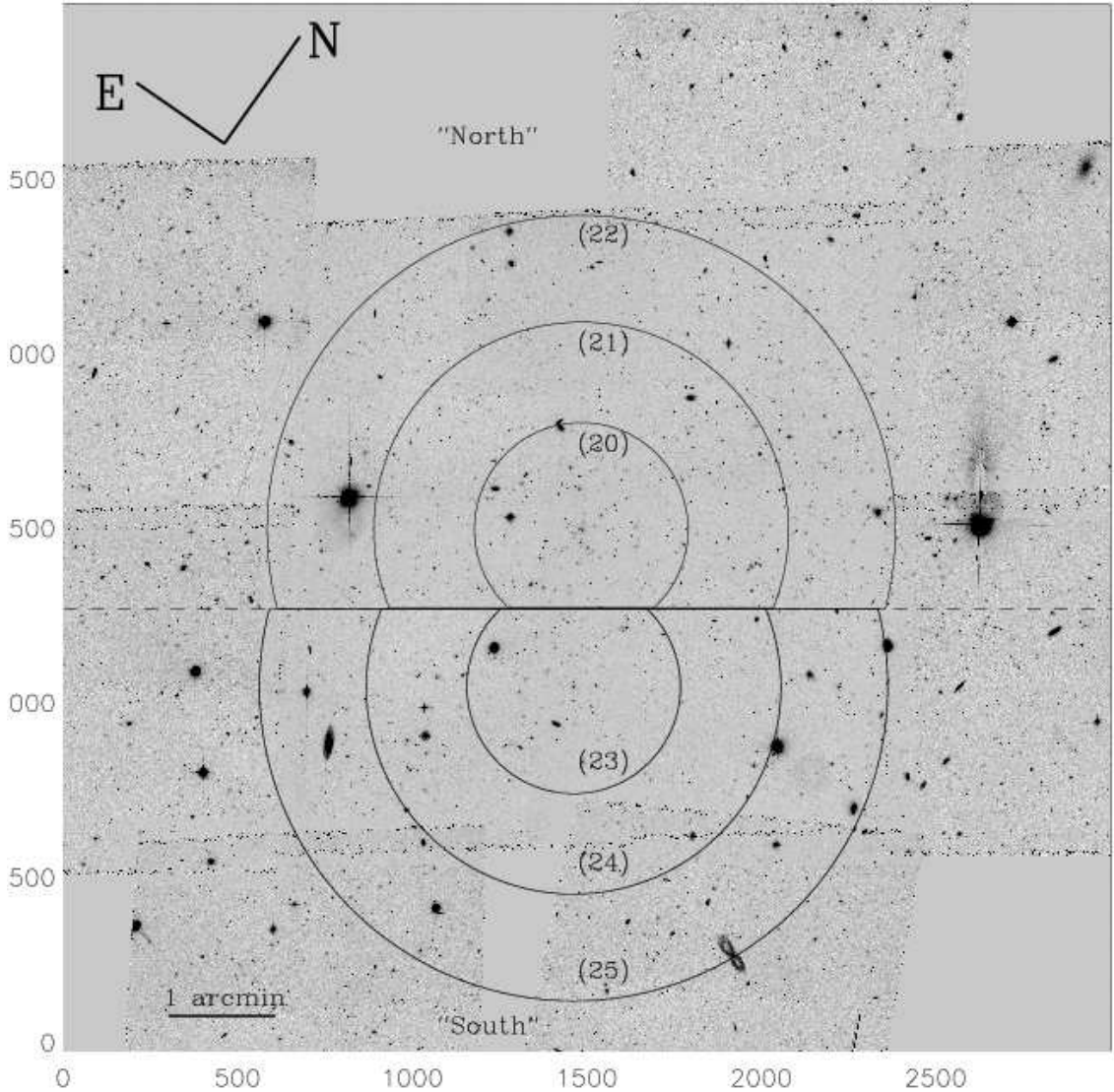}
\caption{Regions used to co-add spectra according to the projected
  angular distribution of the galaxies. Because of the complex
  structure of \CL, composed of two central clumps likely in the
  process of merging, we have defined a set of regions that take into
  account the existence of both sub-clusters. The cluster field of
  view is separated in two halves at a fiducial central point, as
  indicated by the horizontal dashed line. Then we define two groups
  of concentric (semi-)annuli (solid contours) centered at each clump,
  however, truncated at the separation half-way between them. Sectors
  associated with the northern clump (20/N0 through 22/N2) are labeled
  as ``North'', while those associated with the southern clump (23/S0
  through 25/S2) are labeled as ``South''. The background image
  corresponds to the ACS 5\farcm8 $\times$ 5\farcm8 central mosaic,
  complemented by 7 ACS flanking fields subsequently obtained in V
  (F606W) and I (F814W).
\label{rad_reg}}
\end{figure}

We separate the cluster field of view (FoV) in two halves at a
fiducial center (R.A.= 01$h$52$m$41.80$s$,
DEC.=-13$^{o}$57\arcsec52\farcs5) located at the mid-point between the
centers of the northern and southern clumps defined by \citet{drl05},
as indicated by the dashed, horizontal line in Fig. \ref{rad_reg}. We
then consider concentric (semi-)annuli or radial sectors centered at
each clump, however, truncated at the separation half-way between them
(solid contours in Fig. \ref{rad_reg}). Areas or sectors associated with
the northern clump (20/N0 through 22/N2) are labeled as ``North'',
while those associated with the southern clump (23/S0 through 25/S2) are
labeled as ``South''. In table \ref{tab_regs} we give the
corresponding definitions of the above regions. The center of each
clump or subcluster is taken from \citet{drl05}.

All galaxies, passive and those showing emission lines, were
considered for stacking within these regions. We only excluded the
known AGN members.

\subsubsection{Galaxy morphology}

We use the morphological classification given by \citet{pfc05} for
galaxies in \CL. The morphological T-types used by Postman and
collaborators are those defined in \citet{ddc76}. T values ranging
from -5 to -3 correspond to Elliptical (E) galaxies, while a value of
-2 corresponds to S0 galaxies. Values between -1 and 1 are assigned to
morphologies between S0 and Sa, with values $>1$ given to later type
spiral (Sp) galaxies. Type 6 is associated with a Sd morphology, while
irregular (Irr) galaxies have T values in the range $6 < T <
9$. Taking all this under consideration, we established 4 groups in
morphology space as defined in table \ref{tab_regs}. Namely, group
10/E contains elliptical galaxies ($T < -2$); group 11/(S0/Sa),
lenticular galaxies ($-2 \geq T \geq -1$); group 12/Sp, spiral
galaxies ($1 < T \leq 6$); and group 13/Irr, irregular galaxies
($T>6$).

\subsubsection{Stellar mass}

In addition to morphology, we also grouped galaxies based on their
stellar mass content. The work by \citet{hif07} provides stellar mass
estimates for cluster galaxies in \CL\ based on the mass-to-light,
$M/L_B$, ratio and rest-frame $(B-V)$ color linear relation derived by
\citet{bmk03}. Those masses, in despite of being based on a single
color, are proven to be consistent with other estimates as shown by a
comparison with dynamical measurements for some of the same galaxies
(see \citet{hif07}, and references therein).

However, in the analyses shown here, we decided to recompute stellar
masses by using a SED fitting procedure \citep{rrs06} including all
the 5 bands available (see \S\ref{imaging}). This information allowed
us to establish 3 bins in stellar mass (14/RSHM through 16/RSLM) as
defined in table \ref{tab_regs}. Stellar masses span the range
$4.8\times10^9 < M_* \leq 3.9\times10^{11}\ M_{\odot}$, and the mass
interval for each bin has been adjusted in order to have roughly the
same number of spectra to co-add per bin. A more detailed explanation
about the way these stellar masses were obtained is presented in
\S\ref{sed_fit}. In order to study SFH variations with stellar mass
for the same sample of RS galaxies in \S\ref{col_lum_regs}, we also
restrict ourselves to quiescent, RS galaxies, i.e., those with colors
$2.3 < (r_{625}-K_s) < 4.5$ and no visible emission line features.

\subsubsection{Local dark matter density}\label{loc_dmd}

\CL\ has been the subject of a detailed weak lensing analysis by
\citet{jwb05}. By using the available $r_{625}$, $i_{775}$ and
$z_{850}$ ACS data together with photometric and spectroscopic
\citep*[from][]{drl05} redshifts, they are able to measure the shear
signal of the cluster and reconstruct its dimensionless mass density,
$\kappa$. The smoothing scale of the map is $\sim20$\arcsec, while its
accuracy is about 20\%.

As an alternative way of characterizing the cluster environment to
that presented in \S\ref{angdist}, here we use the $\kappa$ map from
\citet{jwb05} to identify environments of different projected mass
density in the ACS FoV of \CL. Because of the so-called sheet-mass
degeneracy, i.e., the invariance of the shear under transformations of
the kind $\kappa \rightarrow \lambda\kappa+(1-\lambda)$, we use this
$\kappa$ map in a relative sense, only, during the interpretation of
the results.

\begin{figure}
\plotone{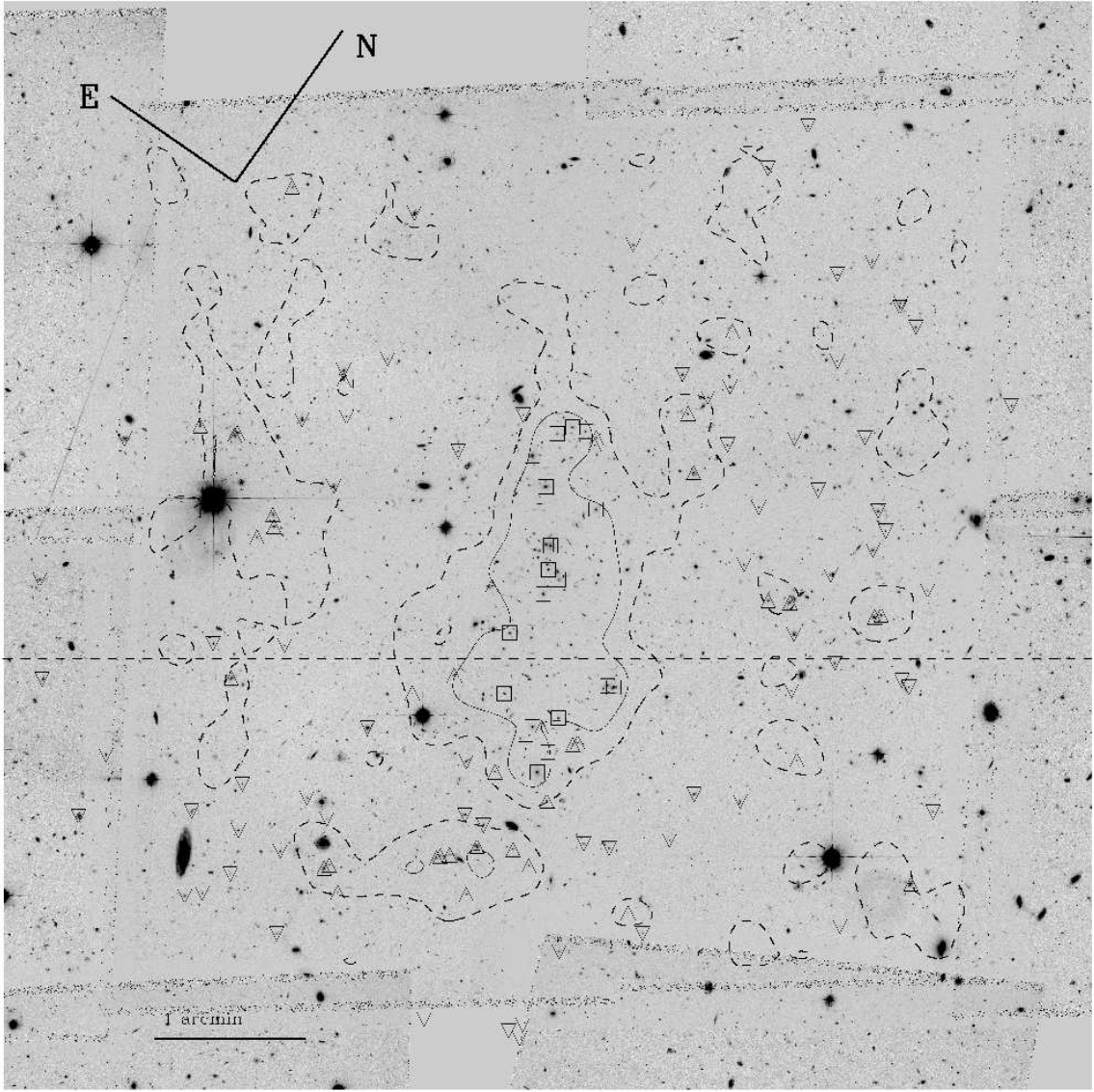}
\caption{Mass density environments defined from the $\kappa$ map of
  \CL\ \citep{jwb05}. Because of the sheet-mass degeneracy, we use
  this map in a relative sense only, during the interpretation of the
  results. The smoothing scale of the map is $\sim20$\arcsec, and its
  accuracy, about 20\%. Three different environments are arbitrarily
  identified. The first of them is characterized by mass densities
  $>20 \times \sigma_{DM}$ (solid contours), with
  $\sigma_{DM}=0.0057\times\Sigma_c$, being $\Sigma_c\sim3650
  M_{\odot}\ pc^{-2}$ the critical mass density of the cluster
  \citep{bhf06}. The $\kappa$ value around the two brightest central
  galaxies of the northern clump \citep*[the cluster center adopted
    in][]{jwb05} is $\sim$0.3. The second one is that containing mass
  densities between 5 (dashed contours) and 20 times $\sigma_{DM}$,
  while the last of the three encompasses mass densities $<
  5\times\sigma_{DM}$, reaching negative values in some areas. These
  three environments correspond to regions 17/HDMD, 18/MDMD, and
  19/LDMD, respectively, as presented in table \ref{tab_regs}. The
  distribution of spectroscopic members is indicated by the
  symbols. Members in the highest density regions are indicated as
  squares; members in the intermediate density regions, as triangles;
  and members in the lowest density environments, as upside-down
  triangles. For comparison, we also show the horizontal dashed line
  in Fig. \ref{rad_reg} that contains the mid-point between the two
  main central sub-clusters. The background image shows the ACS data,
  7\farcm2 a side and centered at the two brightest central galaxies
  of the northern clump.\label{dmd_reg}}
\end{figure}

We thus arbitrarily define 3 different environments based on their
local, projected (total) mass density, as shown in
Fig. \ref{dmd_reg}. The first of them is characterized by mass
densities $>20 \times \sigma_{DM}$ (solid contours in
Fig. \ref{dmd_reg}), with $\sigma_{DM}=0.0057\times\Sigma_c$, being
$\Sigma_c\sim3650 M_{\odot}\ pc^{-2}$ the critical mass density of the
cluster \citep{bhf06}. The $\kappa$ value around the two brightest
central galaxies of the northern clump \citep*[the cluster center
  adopted in][]{jwb05} is $\sim$0.3. The second one is that containing
mass densities between 5 (dashed contours in Fig. \ref{dmd_reg}) and
20 times $\sigma_{DM}$, while the last of the three encompasses mass
densities $< 5\times\sigma_{DM}$, reaching negative values in some
areas. These three environments correspond to regions 17/HDMD,
18/MDMD, and 19/LDMD, respectively, as presented in table
\ref{tab_regs}. Also in Fig. \ref{dmd_reg}, the distribution of
spectroscopic members is indicated by the symbols. Members in the
highest density regions are indicated as squares; members in the    
intermediate density regions, as triangles; and members in the lowest
density environments, as upside down triangles. For comparison, we
also show the horizontal dashed line in Fig. \ref{rad_reg} that
contains the mid-point between the two main central sub-clusters.

As with the galaxies separated according to angular distribution, all
non-emission and emission line galaxies, except the confirmed AGN
members, were considered for stacking. The inclusion here of
star-forming galaxies, as opposed to the regions in color-magnitude
space and stellar mass that only include non-emission line objects, is
because we are also interested in studying how the environment affects
the cluster galaxy population as a whole (passive and star-forming
galaxies; see \S\ref{sfh_environ}). Environmental effects on the
passive cluster galaxy population only is discussed in
\S\ref{sfh_environ} and \S\ref{discussion}.

\subsection{Weighted vs unweighted stacking}\label{w_stack}

After stacking the spectra, we checked for differences between the
co-added spectra obtained from weighting the individual sources by
their corresponding S/N at $\sim4100$ \AA\ (rest-frame; see
\S\ref{stacking}) and those obtained from a direct stacking without
weights. The difference in S/N, $\Delta(S/N)$, as noted in
\S\ref{stacking} and as expected, favors the weight-stacked spectra.

With the exception of a few groups with only 2 to 3 spectra available
for stacking, for which $\Delta(S/N)<1.6$, we note that the region
containing the core of the northern clump (20/N0) has a
$\Delta(S/N)\sim0.5$. The S/N distribution of individual galaxies
within 20/N0 is not specially different from that of many of the other
regions, with a mean S/N of $\sim$ 8 and a
$\sigma_{S/N}\sim3$. Besides this, differences in S/N between the
weight- and unweight-stacked spectra is $\Delta(S/N)>4$, with typical
values of the order of 40\%. Because of the improvement in S/N, the
results that we present below are obtained from the weight-stacked
spectra.

The resulting weight-stacked spectrum for each region in hyperspace is
shown in the corresponding panel of Fig. \ref{specfits_a} and
Fig. \ref{specfits_b} as the thick black line. Each spectrum is shown
redshifted to mean cluster redshift of $z=0.837$ \citep{drl05}. The
relative flux is in arbitrary units.

\subsection{Spectrophotometric fitting}\label{sed_fit}

As in \citet{grs08}, here we perform a fit to both the corresponding
broad-band photometry and the stacked spectrum for each one of the
regions defined in table \ref{tab_regs}. The broad-band magnitude for
a given region of hyperspace in a given band is obtained from
averaging the individual magnitudes of the sources in that region for
that band. The photometric data are weighted by the corresponding
photometric S/N of the object before staking.

In the present analysis, we have used a set of composite stellar
populations models from the \citet{bc03} library, all with solar
metallicity (see \S\ref{metal_range}), and using the ``Padova 1994''
stellar tracks. While a newer, revised library exists \citep{gbb00},
it produces worse agreement with galaxy colors \citep{bc03} than the
Padova 1994 library. Although the \citet{grs08}'s code is able to
handle both the \citet{bc03} and \citet{m05} models, we prefer to use
the former when fitting the co-added spectra because of their higher
resolution ($\Delta \lambda \sim 6$ \AA\ at $z=0.837$) when compared
with the latter ($\Delta \lambda \sim 27$ \AA\ at $z=0.837$), within
the considered wavelength range. In this way, we degrade the
resolution of the redshifted models instead of that of the data.

The model spectra are considered within the context of a delayed,
exponentially declining SFH, without secondary episodes of star
formation \citep*[see][]{grs08}. These SFHs are parameterized by a
time-scale $\tau$, therefore, they are referred to as
``$\tau$-models''. We model the SED as a function of the time $T$
between 200 Myr to the age of the universe at the cluster redshift
($z=0.837$) in increments of $\sim$250 Myr. Different SFHs are
characterized by $\tau$-models with $0 < \tau < 2$ Gyr in increments
of 50 Myr, and we assume a \citet{s55} initial mass function (IMF)
with mass cutoffs at 0.1 and 100 $M_{\odot}$ \citep*[see
  also][]{grs08}. The averaged photometric data and co-added spectra,
for each stacking region, are independently compared with the above
grid of models using a $\chi^2$ statistics with $q$ degrees of freedom
as defined in \citet{a76}.

The number of degrees of freedom, $q$, for each spectrophotometric fit
is defined as the number of independent variables minus the number of
parameters. In the case of the spectroscopic data, the number of
independent variables is difficult to estimate because of correlations
between data points due to a spectral sampling finer than the FWHM
resolution and a linearization of the wavelength solution. In order to
take this into account, and as a good first-order approximation, we
multiplied the $\chi^2$ value of the fit to the stacked spectrum by
the ratio of the FWHM resolution to the sampling size within the
considered wavelength range before estimating the confidence regions.

Since the true SFH of a galaxy may be more complex than a simple
delayed exponential, and because the S/N of galaxy spectra at $z\sim1$
often turns out to be relatively low, the $\tau$-model corresponding
to the absolute minimum $\chi^2$ may not be sufficient to properly
describe the actual SFH. Therefore, and in order to also be less
sensitive to different systematic uncertainties from both the
photometric and spectroscopic data, we consider models that are within
the 99.7\%  confidence regions of the fits to the
photometry and stacked spectra, in the space of model parameters.

Among the models within the intersection of the 99.7\% confidence
regions of the separate fits to the photometric and spectroscopic
data, we consider as ``best-fitting'' model the one which has the
lowest $\chi^2$ value for the fit to the observed co-added
spectrum. In some cases this will be the actual, absolute best fit to
the spectrum, but not always. The best-fitting model in each region of
hyperspace is shown as the blue line in the corresponding panel of
Fig. \ref{specfits_a} and Fig. \ref{specfits_b}. In the same figures,
the minimum and maximum fitting flux values within the above
intersection, and as a function of wavelength, are represented by the
green lines.

The $\chi^2$ statistics \citep{a76} is minimized with respect to the
parameters of the model: $T$, $\tau$, and $M_{*,SED}$, the stellar
mass (for the fit to the photometry), or $\sigma_v$, the stellar
velocity dispersion (for the fit to the spectrum). The value of
$\sigma_v$ is allowed to vary within the range 0 - 400 km/s. We
calculate then the stellar age of a galaxy as its SFR-weighted age,
$T_{SFR}$, defined as:

\begin{equation}
T_{SFR}(T,\tau) = \frac{\int^T_0 (T-t)\psi(t,\tau) dt}{\int^T_0 \psi(t,\tau) dt} \\,
\label{t_sfr}
\end{equation}

\noindent
where 

\begin{equation}
\psi(t,\tau) = \frac{1}{\tau^2} \ t e^{\frac{-t}{\tau}}
\label{tau_mod}
\end{equation}

\noindent
is the $\tau$-model giving the SFR as a function of time since the
onset of the star formation. This definition takes into account the
effective fraction of stellar mass contributed by each single stellar
population making up the model spectrum, and stellar populations
contributing only a negligible fraction to the stellar mass at any
given time $T$ do not affect $T_{SFR}$ significantly. 

We also define a second estimator, the final formation time,
$t_{fin}$, as $M_*(t_{fin})=0.99\times M_*(T)$. In contrast to
$T_{SFR}$, $t_{fin}$ is sensitive to the residual star
formation. Hence, while $T_{SFR}$ measures the age of the bulk of the
stars in a galaxy, $t_{fin}$ traces the last stages of stellar mass
assembly, and is therefore useful to distinguish between two otherwise
old stellar populations that have stopped star formation at different
times. For a model spectrum that fits the observed broad-band
photometry or spectrum of a galaxy, $T-t_{fin}$ corresponds to the
lookback time from the epoch of the galaxy to the last episode of star
formation, and is independent of the time at which the star formation
of the $\tau$-model started.

We note that the results from the spectrophotometric fit are only
weakly dependent on abundance, even though the \citet{bc03} models
used here were computed at solar ratios only. This is due to the fact
that we are also fitting several other spectral features which do not
depend on [$\alpha$/Fe], such as Ca$\mathrm{II}$ H, Ca$\mathrm{II}$ K,
D4000, G4300 \citep{tmb03} and C4668 \citep*[e.g.,][]{jbd05}. The
first three, in particular, are the most prominent features in our
stacked spectra and, therefore, drive the fit on the spectrum. We also
make use of the 5-band SED, whose shape varies with age and
metallicity but does not depend on individual absorption features. As
a result, while the spectral features that do depend on
$\alpha$-abundance might increase the best-fit $\chi^2$, their weight
on the fit is greatly reduced.

In addition, we applied the above $\chi^2$ statistics to the
individual 134 sources spectroscopically confirmed as cluster
members. By fitting their photometric data as described earlier in
this section, we are able to obtain the corresponding stellar mass,
$M_{*,SED}$, which we compare with the value derived by \citet{hif07},
$M_{*,(B-V)_0}$, based on the rest-frame B-V color, if available. Our
photometric stellar masses are about a factor of two larger than those
from \citet{hif07}. A linear fit to the mass measurements gives
$M_{*,SED} = (1.7 \pm 0.27) \times M^{0.99\pm0.3}_{*,(B-V)_0}$
\citep{g09}. The slight overestimate of the SED-derived masses with
respect to those of \citet{hif07} might be due to the choice of the
IMF. Likewise, if the metallicity of cluster galaxies is greater than
solar, their SEDs would appear older when compared with a solar
metallicity model. As a consequence, the fit would tend to
overestimate the near-IR fluxes and thus the stellar masses. We
considered our $M_{*,SED}$ values during the analyses.

\subsection{Spectral features and indices}\label{sp_indx}

In addition to SFHs, we also computed line indices from the stacked
spectra to characterize the stellar and metal content of cluster
galaxies. At rest-frame optical wavelengths, the most prominent
feature in the continuum of a galaxy produced by old, evolved
\citep*[$\gtrsim 3$ Gyr;][]{pb97} stars is the so-called 4000\AA-Break
\citep*[see, e.g.,][]{b83,pb97,e99}. Younger stars, $\lesssim$ 2 Gyr,
have a stronger flux density at wavelengths shorter than 4000 \AA,
producing a different discontinuity: the Balmer jump at 3646
\AA\ \citep*[see][]{e99}.

In addition to the Balmer jump, stars with ages $\gtrsim 0.2$ Gyr also
display large Balmer absorption lines such as H$\alpha$, H$\beta$,
H$\gamma$, H$\delta$, H$\epsilon$, H6,
etc. \citep*[e.g.,][]{sbb05}. The strength of these lines is maximum
for A0V stars, but they can be detected from late-B to early-F type
stars \citep{pb97}. The presence of deep Balmer lines is the signature
of a young stellar population \citep*[e.g.,][]{cs87,psd99}.

As the stellar population ages, the main contribution to the flux
shifts to cooler stars. The luminosity of the galaxy decreases, as
does the depth of the Balmer lines, and the Balmer jump is replaced
with the 4000 \AA-break produced by line blanketing due to metals
including CN and Ca$\mathrm{II}$ \citep*[e.g.,][]{e99}. The strength
of the 4000 \AA-break increases with age and metal content and is, for
a fixed metallicity, a measure of age \citep*[e.g.,][]{pb97,khw03} and
a good indicator of old populations of stars.

Because of the aforementioned metal absorption, the true continuum at
some wavelengths cannot be measured, and the apparent strength of some
of the high-order Balmer features (H$\delta$, H$\epsilon$, H6, etc.)
becomes sensitive to the metal content and [$\alpha$/Fe] ratio of the
stellar population \citep{mgr03,dop04,tmk04,prc07}. Therefore, line
indices based on such features have to be considered with caution when
using them to quantify young stellar populations in galaxies. Another
effect that may bias this quantification is that of the intrinsic
velocity dispersion of the galaxy, however, of smaller magnitude
\citep*{kif06} compared with that of metallicity.
  
Very young ($< 200$ Myr), massive ($> 10 M_{\odot}$) O- and B-type
stars are able to ionize their surrounding gaseous medium, which
translates in the presence of emission line features such as
H$\alpha$, H$\beta$, [O$\mathrm{II}$] ($\lambda$$\lambda$3727) and
[O$\mathrm{III}$] ($\lambda$4959,$\lambda$5007) in the optical
window. Because of the short lifetimes of those massive, ionizing
stars, these emission lines can be used to obtain a measure of the
nearly instantaneus SFR, independent of the previous star formation
history \citep{k98}.

Establishing the significance of the old stellar content of a galaxy
is done my means of the D4000 index \citep{b83} that measures the
ratio between the continuum level at both sides of the 4000
\AA-break. Here we use the modified version of it \citep*[][see table
  \ref{tab_ind_def}]{bmy99}, which is based on narrower continuum
windows\footnote{Also referred to as D$_n$(4000); see
  \citet{khw03}.}. Although sensitive to metallicity effects, at early
stages ($\lesssim 1$ Gyr) of stellar evolution, the D4000 index can be
used as a good age indicator \citep{pb97,khw03}. Assuming solar
metallicity, stellar populations older than 3 Gyr are characterized by
D4000 values $>1.7$ \citep{khw03}.

The significance of young stars, between 1 and 2 Gyr old, is estimated
from the equivalent width (EW) of some of the Balmer absorption lines
\citep*[see, e.g.,][]{cs87,psd99}. Here we use the H$\delta_A$ index,
as defined by \citet{wo97}. Following \citet{dsp99}, we consider
objects with EW(H$\delta_A$) $>4$ \AA\ \citep*[see also][]{psd99} as
galaxies with a significant young stellar component. For younger ages,
the [O$\mathrm{II}$] index defined in \citet{tfi03} (see table
\ref{tab_ind_def}) can be used as an indicator of current star-forming
activity, keeping in mind its vulnerability to dust absorption. As in
\citet{dsp99}, we consider values of EW([O$\mathrm{II}$]) $\lesssim
-5$ \AA\ as significant. As shown recently for a supercluster
environment at $z\sim0.9$ \citep{lls10}, [O$\mathrm{II}$] emission can
also be associated with a LINER or Seyfert component. Although it is
something to bear in mind, we assume that this is not the case for our
[O$\mathrm{II}$] emitters.

In addition to H$\delta_A$, we introduce a new index associated with
the H6 ($\lambda$3889) Balmer line \citep*[e.g.,][]{vs03}. The
pseudo-continuum and line windows are given in table
\ref{tab_ind_def}. We use a single stellar population, solar
metallicity, 0.5 Gyr old model from the Bruzual-Charlot library
\citep{bc03} in order to define the windows used to calculate the line
EW. We note, however, that this definition becomes sensitive to the
strength of CN when old stellar populations become more important in
the galaxy spectrum.

Finally, in order to investigate trends with metallicity, we also
computed indices of some metal features available in the wevelength
range covered by our data: Fe4383 and C4668, as defined in
\citet{wfg94}, and CN3883, as defined in \citet{dc94} (see table
\ref{tab_ind_def}).

Most of the absorption indices described above are measured here using
the same window definitions of the Lick/IDS system
\citep{wfg94,wo97}. However, our spectra have a resolution that is
$\sim 3-5$ \AA\ lower than those in the Lick/IDS system \citep{wo97},
therefore, a direct comparison to the Lick/IDS indices cannot be
performed.

\section{Results}\label{results}

\subsection{Star Formation Histories}\label{sfh}

\begin{figure}
\plotone{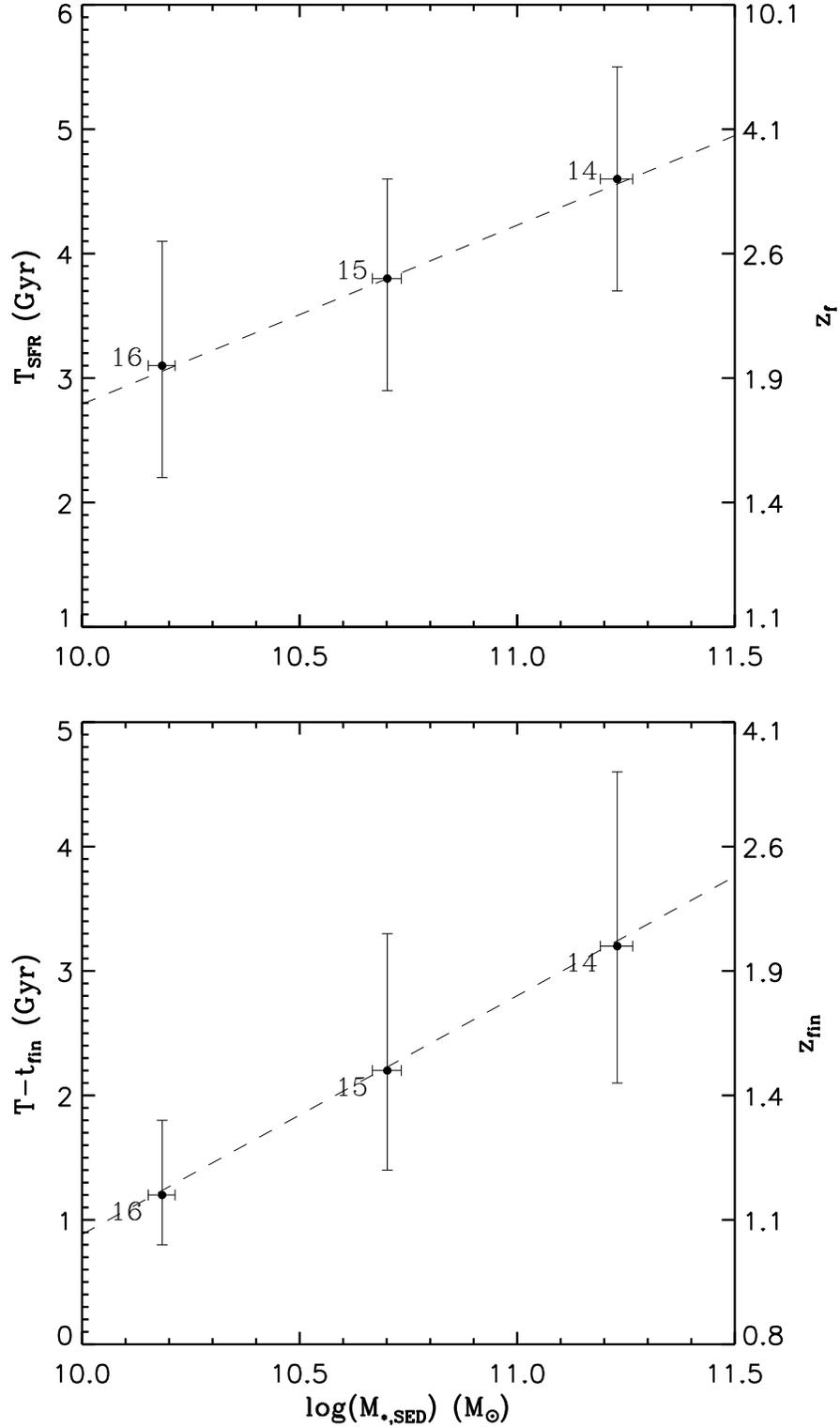}
\caption{SFHs of passive RS galaxies as a function of average stellar
  mass, $M_{*,SED}$, for the three mass bins, 14/RSHM, 15/RSMM and
  16/RSLM. Average stellar mass is obtained by averaging the
  individual stellar masses of galaxies in each bin, and the
  corresponding error bars are calculated as the standard error. The
  dashed lines represent linear fits to the data taking into account
  error bars. Lower stellar mass bins tend to have, on average,
  younger ages and more extended periods of active star formation (see
  \S\ref{sfh_m_rs} for details).\label{sfh_mass}}
\end{figure}

\begin{figure}
\plotone{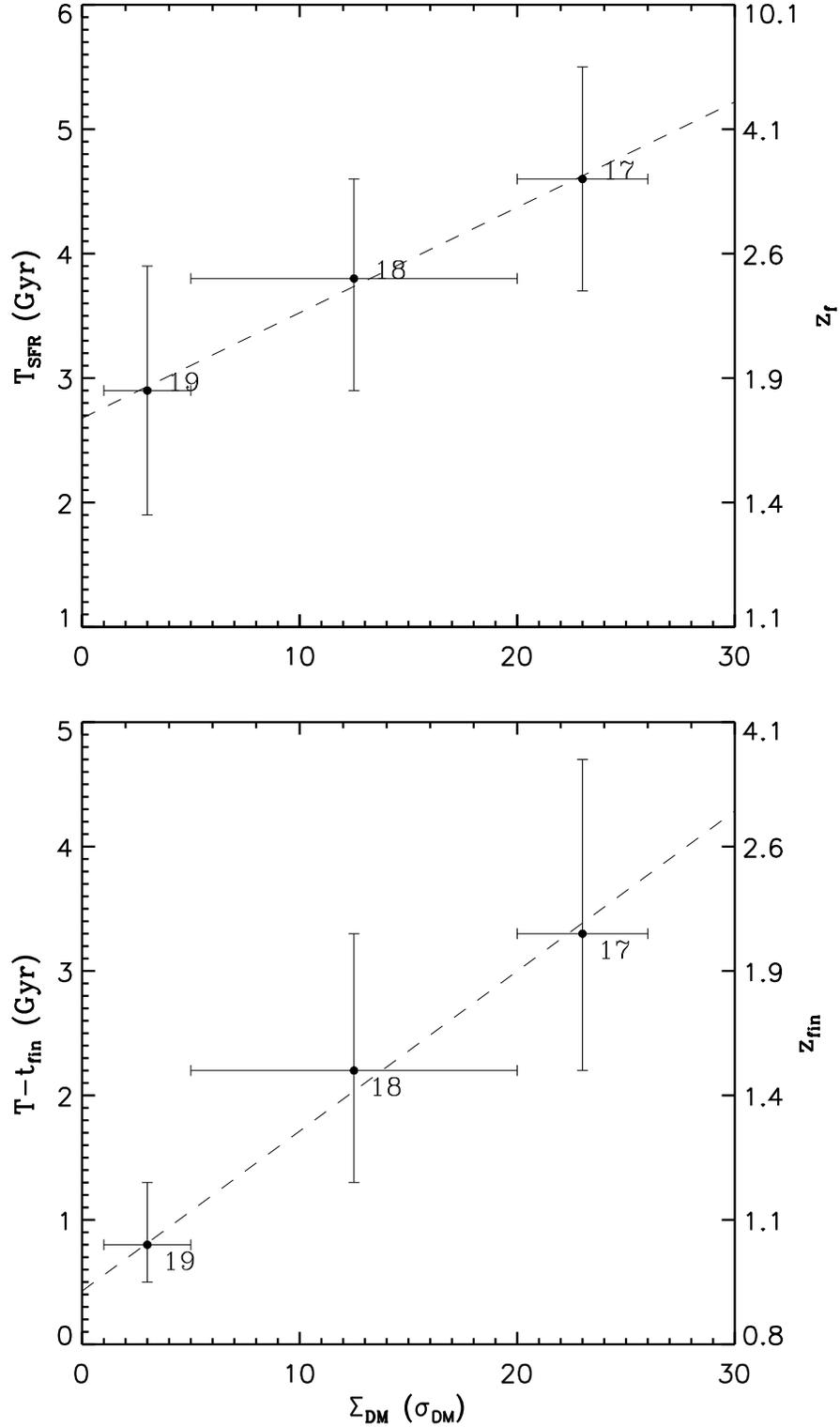}
\caption{SFHs of cluster galaxies as a function of local projected DM
  density, $\Sigma_{DM}$, for the three bins, 17/HDMD through
  19/LDMD. The dashed lines represent linear fits to the data taking
  into account error bars. The results are consistent with each other
  within the errors. On average, the bulk of the stars in cluster
  galaxies were formed at roughly the same time, $z_f\sim2.6$ which
  corresponds to a $T_{SFR}\sim3.8$ Gyr. However, galaxies in the
  lowest DM density environment (19/LDMD) form stars down to $\sim0.8$
  Gyr prior to the epoch of observation ($z=0.837$), while the average
  galaxy in the highest density region (17/HDMD) stopped forming stars
  $\sim3.2$ Gyr prior to the epoch of observation (see
  \S\ref{sfh_environ}).\label{sfh_env}}
\end{figure}

\begin{figure}
\plotone{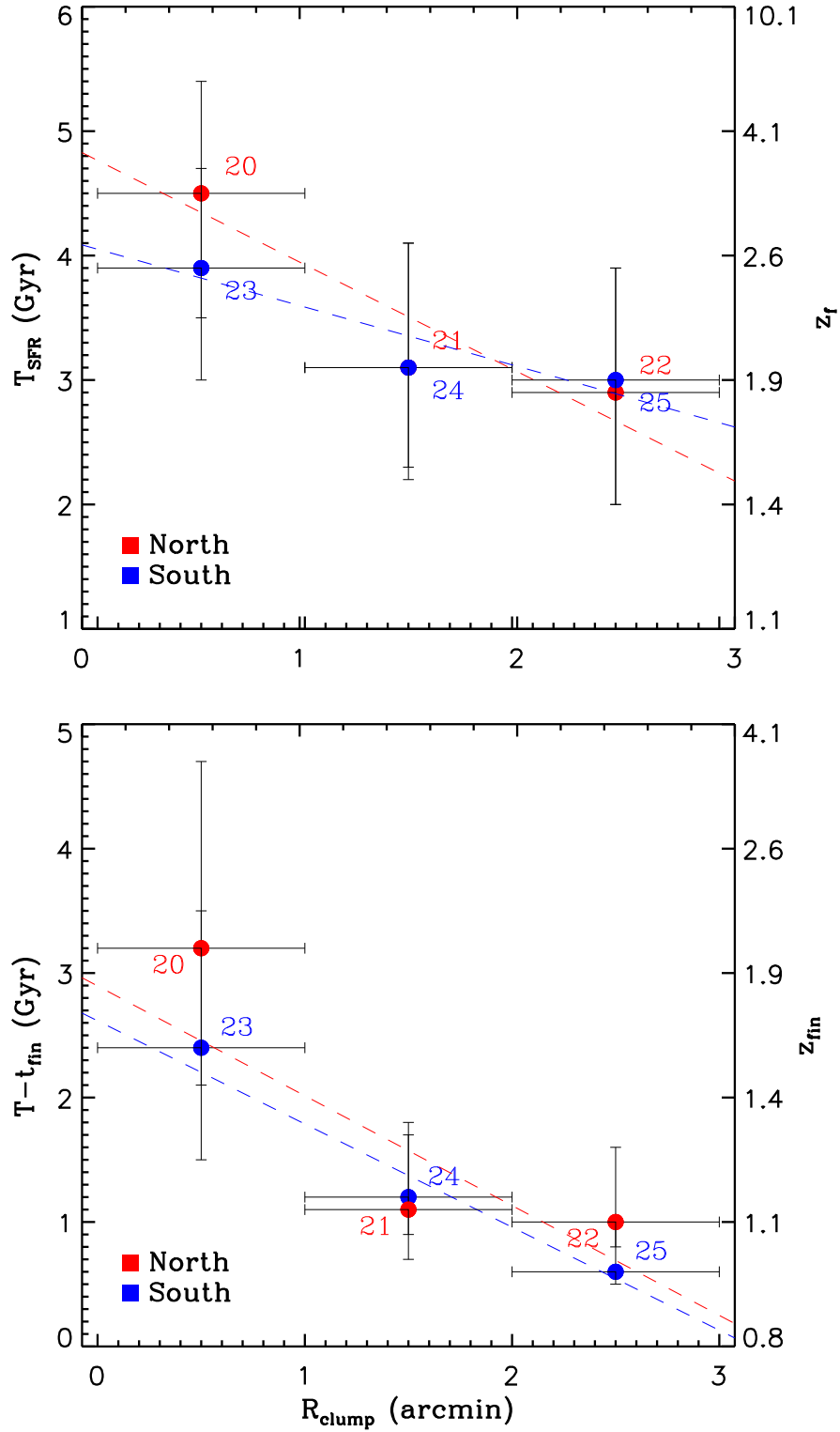}
\caption{SFHs of cluster galaxies as a function of radial distance
  from the center of each of the two main subclusters, as defined in
  table \ref{tab_regs}. Red points correspond to regions 20/N0 through
  22/N2 in the northern subcluster while blue points correspond to
  regions 23/S0 through 25/S2 in the southern subcluster. The
  corresponding linear fits are shown as dashed red and blue
  lines. Although not statistically significant, there is some
  indication that the central region of the northern subcluster is
  slightly older and has stopped forming stars earlier than the
  central region of the southern subcluster (see
  \S\ref{sfh_environ}).\label{sfh_rad_sect}}
\end{figure}

The spectrophotometric fitting procedure described in \S\ref{sed_fit}
allows us to characterize the SFH associated with the co-added
photometry and spectra of the galaxies in each of the hyperspace
regions defined in table \ref{tab_regs}. This characterization is
given in terms of the SFR-weighted age, ($T_{SFR}$; see
Eq. \ref{t_sfr}), the formation redshift ($z_{f}$), the final
formation lookback time from $z=0.837$ ($T-t_{fin}$; see
\S\ref{sed_fit}), and the final formation redshift ($z_{fin}$; see
\S\ref{sed_fit}). The formation redshift, $z_f$, is defined as the
redshift corresponding to $T_{SFR}$. The values of these parameters
for the different regions defined in table \ref{tab_regs} (see
\S\ref{indx_regs}) are summarized in table \ref{tab_sfh}. ``Error''
bars shown in Figures \ref{sfh_redseq_cm} through \ref{sfh_rad_sect}
actually indicate the maximum and minimum parameter values within the
intersection of the 99.7\% confidence regions of the fits to the
composite SED and stacked spectrum.

When fitting the spectrophotometric data of bins that have a mix of
early- and late-type galaxies, the adopted SFH may not be sufficient.
Indeed, the spectrophotometry of those bins corresponds to a
combination of old and young stellar populations, whose composite SFH
may not be adequately parametrized by a simple delayed exponential. In
addition, the [O$\mathrm{II}$] ($\lambda$3727) emission feature is
enclosed by the $r_{625}$ band. This results in a higher $r_{625}$
flux and therefore in bluer colors. The best fitting models to the SED
would then be younger than those to the spectrum, where the
[O$\mathrm{II}$] line is ignored. This can lead to a significant
discrepancy between the photometric and spectroscopic solutions, with
no intersection in parameter space. In fact, no 99.7\% intersection
in parameter space was found for region 24/S1 (see table
\ref{tab_sfh}).

In what follows, we present our main results about the average SFH of
cluster members in the RS, as well as in terms of stellar mass and
environment, as defined in \S\ref{indx_regs} (see table
\ref{tab_regs}). Since SFHs are obtained by using solar metallicity
models, in \S\ref{metal_range} we give a justification for this while
in \S\ref{metal_effects} we discuss about the biases introduced by
this choice.

\subsubsection{SFH in the cluster RS}\label{sfh_rs}

\begin{figure}
\plotone{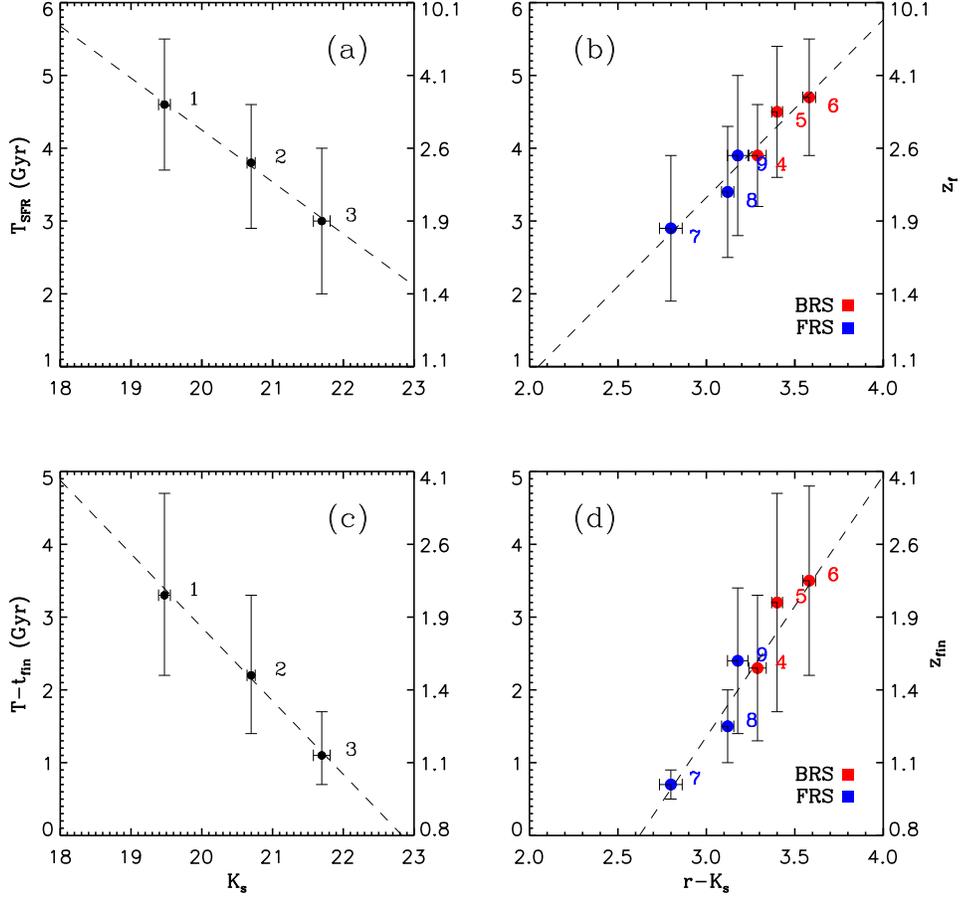}
\caption{SFHs for galaxies in the RS as a function of $K_s$ brightness
  and $r_{625}- K_s$ color. Results from the three-bin partition
  (regions 1/BRS through 3/FRS) are shown in panels (a) and
  (c). Results from the finer RS partition (regions 4/BBRS through
  9/FRRS) are shown in panels (b) and (d). For clarity, bins in the
  bright half of the red-sequence (BRS) and in the faint half of the
  red-sequence (FRS) are shown in red and blue, respectively. The
  dashed lines are the best linear fits to the data points considering
  error bars. On average, galaxies in the brighter and redder RS bins
  are older and also display shorter periods of active star formation,
  i.e., longer $T-t_{fin}$ lookback times (see \S\ref{sfh_rs} for
  details).\label{sfh_redseq_cm}}
\end{figure}

Results from the three-bin partition are shown in panels (a) and (c)
of Fig. \ref{sfh_redseq_cm}. Panel (a) shows that the $T_{SFR}$ within
the cluster RS decreases proportional to $K_s$ as $T_{SFR} = -0.7
\times K_s + 18.5$ Gyr. Panel (c) shows that the lookback time from
the epoch of observation to the last episode of star formation,
$T-t_{fin}$, varies as $T-t_{fin} = -1.0 \times K_s + 23.0$ Gyr. The
other two panels in Fig. \ref{sfh_redseq_cm}, (b) and (d), show the
variations of $T_{SFR}$ and $T-t_{fin}$ with color as obtained from
the finer RS partition. The Spearman's rank test \citep{ptv92}
indicates that the trend between age and $r_{625} - K_s$ color (panel
b) is significant with $>95$\% significance ($\rho=0.99$). This
correlation can be described as $T_{SFR} = 2.4 \times (r_{625} - K_s)
- 4.0$ Gyr. In the case of the lookback time $T-t_{fin}$, the
Spearman's rank test also gives a significance $>95$ \% ($\rho=0.94$)
for the observed correlation with color. A linear fit shows that this
correlation can be characterized as $T-t_{fin} = 3.6 \times (r_{625} -
K_s) - 9.4$ Gyr.

The above correlations indicate that, on average, galaxies in the
brighter and redder RS bins are older and also display shorter periods
of active star formation. Galaxies at the bright-end of the RS would
have formed the bulk of their stars at $z_f\gtrsim3$ and finished
their star formation at $z_{fin}\sim2$, while those in the faint-end
of the RS would have formed most of their stars at $z_f\sim2$. This
formation redshift for the average galaxy at the bright-end of the RS
is in agreement with the estimates by \citet{bhf06} for cluster
elliptical galaxies in \CL, and with recent measurement of $z_f$ on
brightest cluster galaxies at $z>1.2$ \citep*[][see also Papovich et
  al. 2010]{csh09}. In particular, while regions 8/FGRS and 9/FRRS
have similar SFR-weighted ages and final formation times, the
faint-blue end of the RS (7/FBRS) is very young, with a final
formation redshift of $z_{fin}\sim1$, less than 1 Gyr from the epoch
of the cluster.

A similar spread toward younger ages in faint, low-mass galaxies is
reported by \citet{gcb06} from their SDSS sample of early-type
galaxies. In the case of the three bright bins, 4/BBRS thourgh 6/BRRS,
we observe clear trends in the sense that both $T_{SFR}$ and the
$T-t_{fin}$ loockback time increase with $(r_{625} - K_s)$ color. All
these correlations with color, in the bright and faint parts of the
RS, support the conclusion that the intrinsic scatter of the RS is
mainly due to stellar age differences.

\subsubsection{SFH as a function of stellar mass in the cluster RS}\label{sfh_m_rs}

Considering only passive RS members, we find that both $T_{SFR}$ and
$T-t_{fin}$ scale with average stellar mass, $M_{*,SED}$, as shown in
Fig. \ref{sfh_mass}. The average mass in each mass bin (14/RSHM,
15/RSMM and 16/RSLM) is obtained by averaging the individual stellar
masses of galaxies in each bin, and the corresponding error bars are
calculated as the standard error. We find that lower stellar mass bins
tend to have, on average, younger ages and more extended periods of
active star formation. Linear fits to the data give $T_{SFR}= 1.4
\times log(M_{*,SED}) -11.6$ Gyr, and $T-t_{fin}= 1.9 \times
log(M_{*,SED}) - 18.3$ Gyr, in qualitative agreement with the results
from the fit to the color-luminosity selected bins. This is not
surprising as the K$_s$-band luminosity is a good tracer of the
stellar mass.

The average galaxy population in the most massive bin (14/RSHM) is
characterized by having formed the bulk of its stars at $z_f\gtrsim3$
and stopped its star-forming activity by $z_{fin}\sim2$. On the other
hand, the average SFH from the best-fitting models to the co-added
spectrophotometric data of the lowest mass bin (16/RSLM) shows a delay
of $\sim2.2$ Gyr, with a formation redshift $z_f\sim2$ and a final
formation redshift $z_{fin}\sim1$.

These results show that more massive galaxies stopped forming stars
earlier, in this case $\sim2$ Gyr, than less massive ones, consistent
with the ``downsizing'' scenario for the star formation in galaxies
proposed by \citet{csh96}. As the average stellar mass of passive
cluster members in the faint-blue RS bin (7/FBRS) is consistent with
that of the other two faint RS bins, 8/FGRS and 9/FRRS (see table
\ref{tab_sfh}), we conclude that the age difference found for region
7/FBRS with respect to the other RS regions is not (solely) due to
stellar mass. Hence, non-intrinsic factors such as the cluster
environment must be taken into account.

\subsubsection{SFH as a function of environment}\label{sfh_environ}

The star formation-weighted age and the $T-t_{fin}$ lookback time as a
function of local projected DM density, i.e., regions 17/HDMD through
19/LDMD, are shown in Fig. \ref{sfh_env}. The results are consistent
with each other within the parameter ``errors'' (see \S\ref{sfh}). On
average, the bulk of the stars in cluster galaxies were formed at
roughly the same time, $z_f\sim2.6$ which corresponds to a
$T_{SFR}\sim3.8$ Gyr (see table \ref{tab_sfh}). Linear fits to the
data give $T_{SFR}=0.1 \times \Sigma_{DM} + 2.7$ Gyr and
$T-t_{fin}=0.1 \times \Sigma_{DM} + 0.5$ Gyr.

However, the signature of the local environment can clearly be seen:
galaxies in the lowest DM density environment (19/LDMD) form stars
down to $z_{fin}\sim1$, that is $\sim0.8$ Gyr prior to the epoch of
observation ($z=0.837$). In contrast, the average galaxy in the
highest density region (17/HDMD) has stopped forming stars $\sim3.2$
Gyr prior to the epoch of observation ($z_{fin}\sim2$).

Not surprisingly, this increase (decrease) in duration of the star
formation activity ($T-t_{fin}$) when going to lower projected DM
density regions is qualitatively consistent with the variations in
$T-t_{fin}$ when moving from central to external areas of each
subcluster. These trends are shown in Fig. \ref{sfh_rad_sect}. Red
points correspond to regions 20/N0 through 22/N2 in the northern
subcluster while blue points correspond to regions 23/S0 through 25/S2
in the southern subcluster.

The corresponding linear fits are shown as dashed red and blue
lines. These fits are described as $T_{SFR}=-0.8 \times d_{clump} +
4.8$ Gyr and $T-t_{fin}=-0.8 \times d_{clump} + 2.6$ Gyr for the
northern radial sectors, and as $T_{SFR}=-0.5 \times d_{clump} + 4.1$
Gyr and $T-t_{fin}=-0.8 \times d_{clump} + 2.6$ Gyr for the southern
radial sectors. $d_{clump}$ corresponds to the distance between the
center of the subcluster and the radial mid-point of a given radial
sector (see Fig. \ref{rad_reg} and table \ref{tab_regs}).

There is also some indication that the central region of the northern
subcluster is slightly older and has stopped forming stars earlier
than the central region of the southern subcluster.

We point out that most of the galaxies considered in region 7/FBRS are
located in the outskirts of \CL, within the low DM density bin
corresponding to bin 19/LDMD. Therefore, it is likely that the younger
age derived from the stacked spectrophotometric data of 7/FBRS is
related to the local environment. We will discuss this in more detail
later on in \S\ref{discussion}. It is important to say that this
dependence of SFH with environment is consistent with the
environmental dependence of galaxy colors found by \citet{bhf06} in
\CL, and of star formation found in low redshift high-density
environments \citep{gwm04}.

\begin{figure}
\plottwo{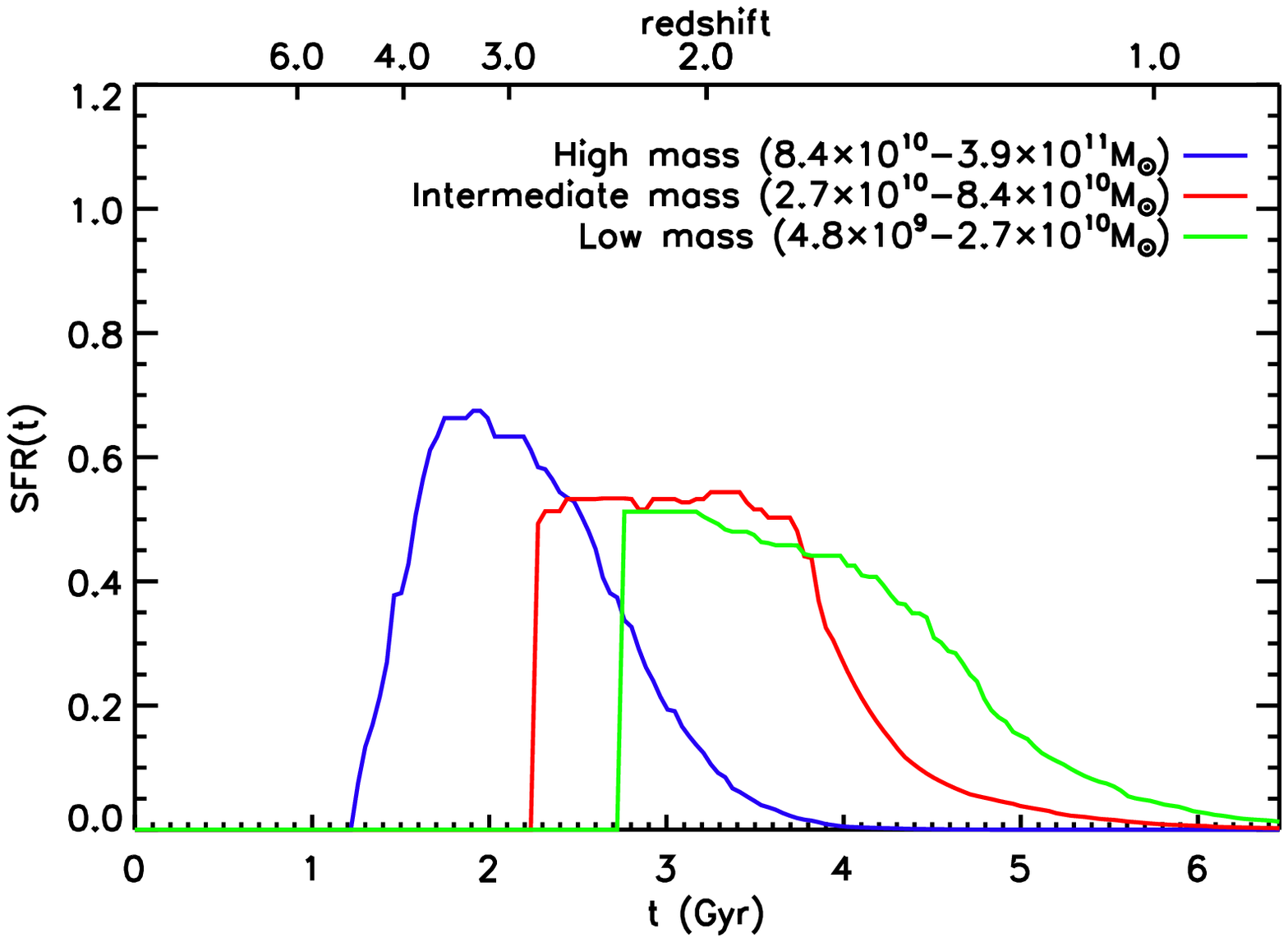}{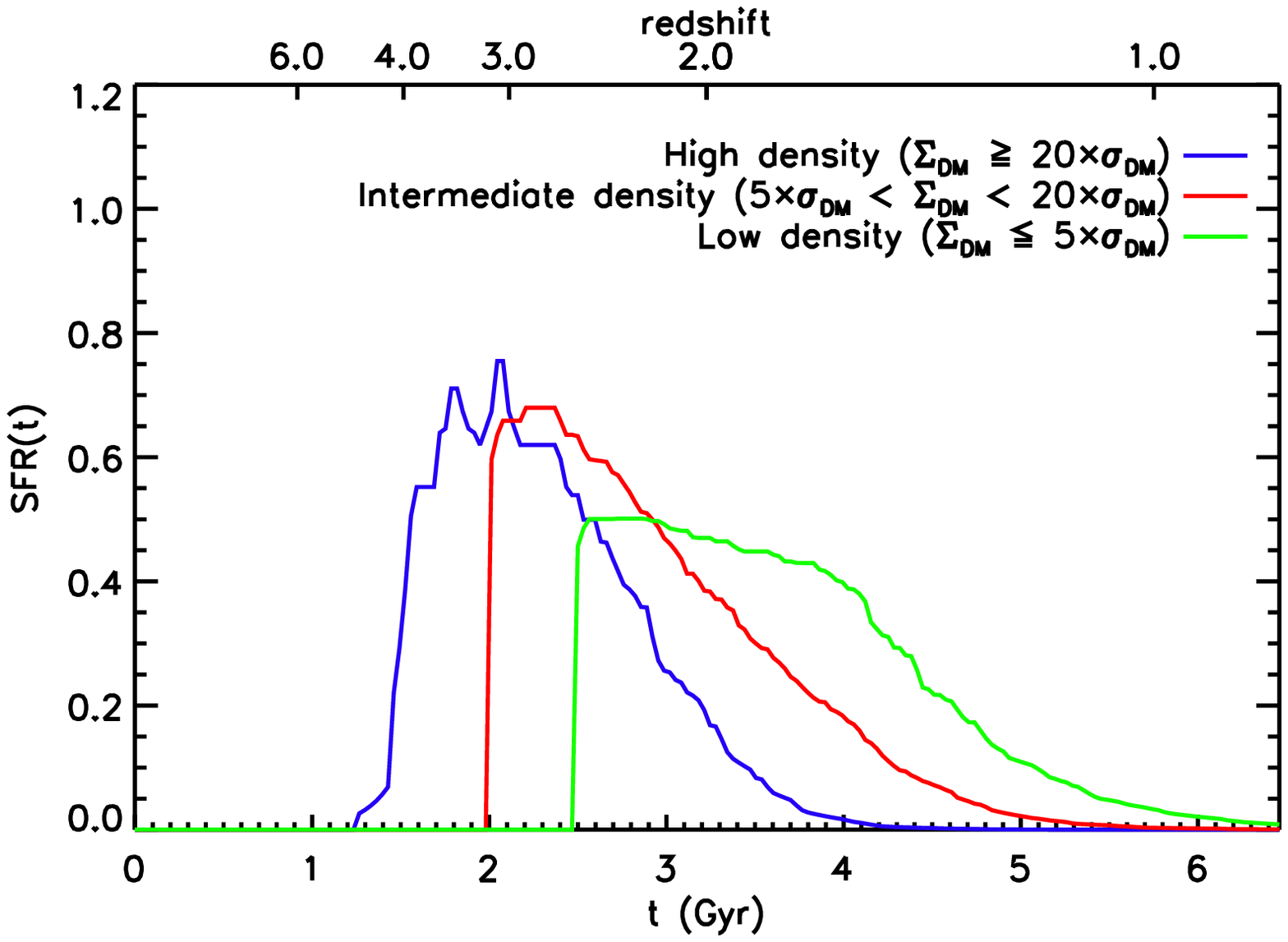}
\caption{Median SFHs of the best-fitting models of the
  spectrophotometric data of non-[O$\mathrm{II}$] galaxies in the
  stellar mass-selected regions, 14/RSHM through 16/RSLM ({\it left}),
  and local dark matter density regions, 17/HDMD through 19/LDMD ({\it
    right}). The bottom axis shows the cosmic time $t$ and the top
  axis shows the corresponding redshift. High mass galaxies and those
  in the highest density environments have formed the bulk of their
  stars at $z>3$ and stopped their star-forming activity at $z\sim2$
  altogteher, with the most massive galaxies ($> 8\times10^{10}
  \ M_{\odot}$) being $\sim1$ Gyr older than the less massive ones
  ($3\times10^{10} \ M_{\odot}$). These ages suggest a formation
  scenario involving an accelerated SFH and early quenching of star
  formation.\label{sfh_mass_dens}}
\end{figure}

If we concentrate only on the cluster members that do not show
[O$\mathrm{II}$], the corresponding SFHs are observed to be different
depending on stellar mass and local environment. This is illustrated
in Fig. \ref{sfh_mass_dens}. In it we show the SFHs of the
best-fitting models of the spectrophotometric data in the stellar
mass-selected regions of RS galaxies, 14/RSHM through 16/RSLM, and
local dark matter density regions, 17/HDMD through 19/LDMD. The figure
indicates that high mass galaxies and those in the highest density
environments have formed the bulk of their stars at $z>3$ and stopped
their star-forming activity at $z\sim2$ altogteher, with the most
massive galaxies ($> 8\times10^{10} \ M_{\odot}$) being $\sim1$ Gyr
older than the less massive ones ($3\times10^{10} \ M_{\odot}$). These
ages suggest a formation scenario involving an accelerated SFH and
early quenching of star formation, possibly followed by further mass
assembly via mergers (see \S\ref{discussion}).

\subsection{Spectral indices}\label{res_indx}

Spectral indices were obtained directly from the stacked spectra of
the different regions in the hyperspace defined in
\S\ref{indx_regs}. Our results are presented below.

\subsubsection{The red-sequence}\label{spind_rs}

\begin{figure}
\plotone{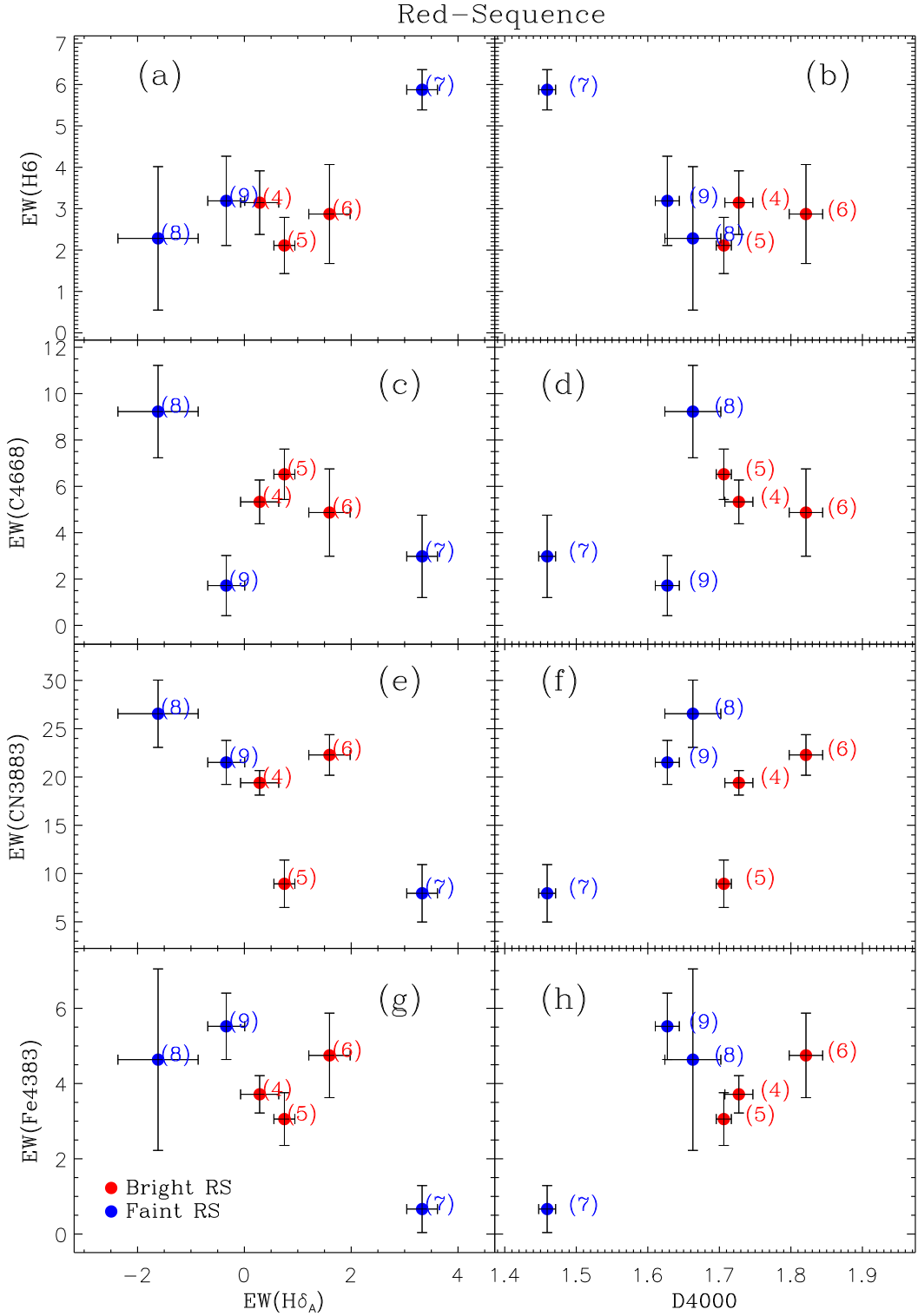}
\caption{Distribution of spectral indices for galaxies in the
  red-sequence. Red circles correspond to regions in the bright half
  of the RS while blue circles correspond to regions in the faint
  half. Only passive galaxies, i.e., galaxies with no detectable
  [O$\mathrm{II}$] in emission, have been stacked in each region. In
  general, notable deviations ($>2\sigma$) of region 7/FBRS with
  respect to the brightest and reddest region (6/BRRS) in the
  red-sequence can be seen for most of the indices
  shown.\label{redseq}}
\end{figure}

Fig. \ref{redseq} shows line strengths for the 6 bins in the cluster
RS (see table \ref{tab_regs}). Red circles correspond to to regions in
the bright half of the RS while blue circles correspond to regions in
the faint half. Only passive galaxies, i.e., galaxies with no
detectable [O$\mathrm{II}$] in emission, have been stacked in each
region.

We find that the central and red bins of the faint half of the RS
(reions 8/FGRS and 9/FRRS), as well as the bright half of the RS
(regions 4/BBRS through 6/BRRS), show little to no H$\delta$
absorption with a pronounced 4000\AA-break, consistent with an old
$>2$ Gyr old population \citep{pb97}. In contrast, the stacked
spectrum corresponding to the blue faint-end of the RS (region 7/FBRS)
shows a moderate H$\delta$ absorption (EW(H$\delta_A$)$>3$), falling
in the $k+a$ category of \citet{dsp99}, and a strong ($\sim6$ \AA) H6
index. At the same time, the D4000 index of 7/FBRS is the weakest
among our measurements in the RS. Thus, the most notable feature of
these diagrams is the separation of the blue, faint-end of the RS
(region 7/FBRS) from the rest of the other RS regions for those
age-sensitive indices such as D4000, H$\delta_A$ and H6.

The composite spectrum of the blue faint-end of the RS is thus
consistent with that of a quiescent stellar population which
experienced its latest episode of star formation about 1.5 Gyr
earlier, i.e., that of a post-star-forming galaxy
\citep*[][]{cs87,psd99}. This suggests that, while the bright half of
the RS appears fully assembled at $z = 0.837$, the blue faint-end of
the RS is still in the process of being populated via the migration of
low-mass ($\lesssim5 \times 10^{10} M_{\odot}$) galaxies from the blue
cloud as their star formation is suppressed. At this redshift, this
mass limit is broadly consistent with the transition mass found for
field galaxies using the luminosity functions of both early-type and
star-forming galaxies \citep*[e.g.,][]{cdr06,bec06}.

In terms of metal indices, the above separation between bin 7/FBRS and
the rest of the RS becomes less obvious for indices such as C4668 and
CN3883, although it can be seen for the Fe4383 index. In general,
except for C4668, a deviation of at least 2$\sigma$ is measured
between region 7/FBRS and the brightest, reddest region (6/BRRS) in
the RS for most of the indices shown. This segragation is likely the
manifestation of significant differences in age and metal content of
galaxies at opposite ends of the RS.

\subsubsection{The cluster environment}

\begin{figure}
\plotone{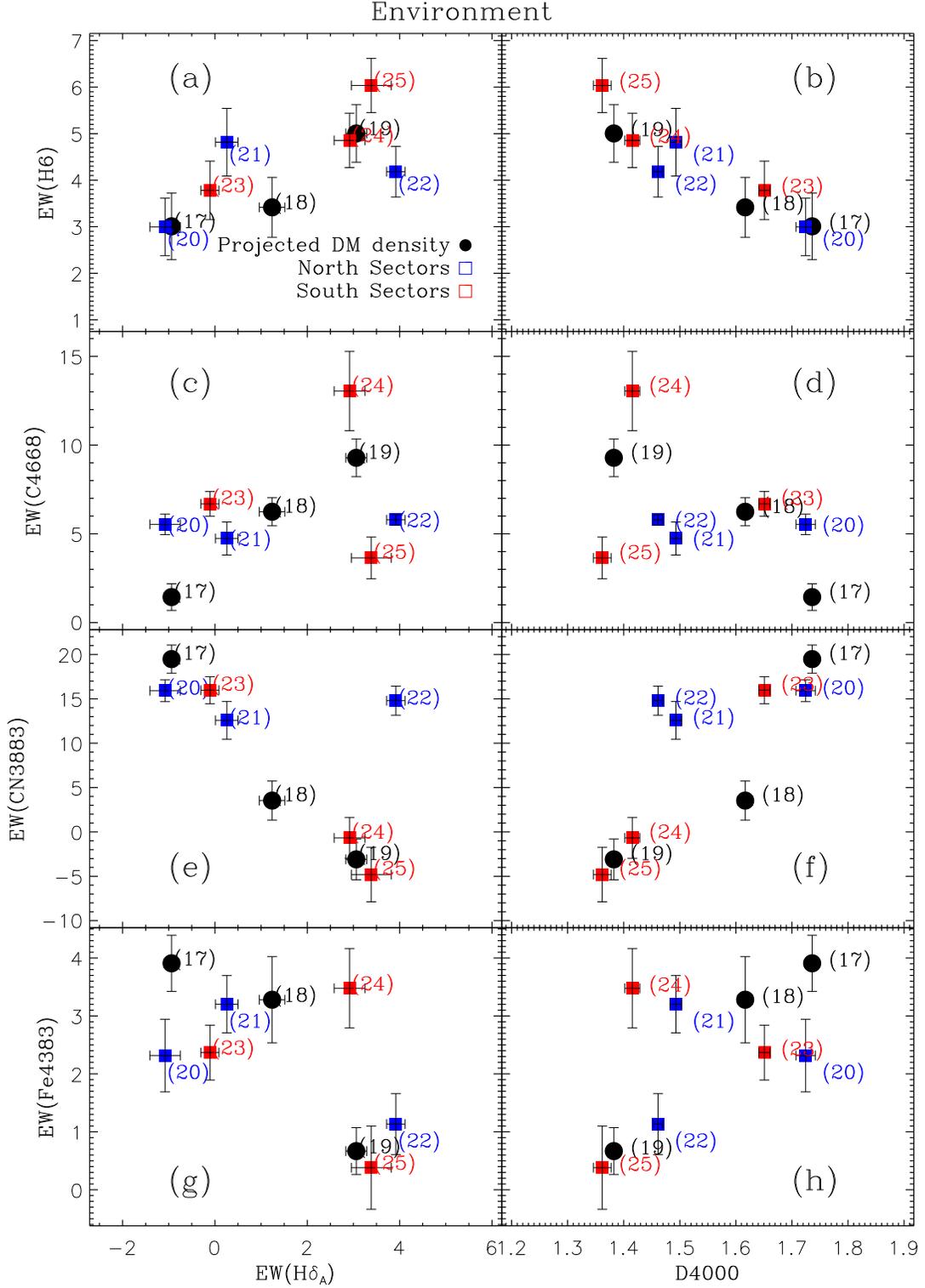}
\caption{Distribution of spectral indices for galaxies grouped
  according to local environment. The latter is characterized in three
  different ways: projected DM density (black circles), angular
  distribution in the northern subcluster (blue squares), and angular
  distribution in the southern subcluster (red squares), as defined in
  \S\ref{indx_regs} (see table \ref{tab_regs}). All (passive and
  star-forming, but no AGN) galaxies have been stacked in each
  region. Clear trends of the age-sensitive indices with environment
  are observed. H$\delta_A$ and H6 increase, on average, toward the
  outskirts of the subclusters and with decreasing projected DM
  density, while the opposite trend is observed for the D4000
  index. As opposed to the EW(C4668), the EW(CN3883) and EW(Fe4383)
  show a notable decrease toward lower DM density regions. Gradients
  with radial distance are only notable for the CN3883 and Fe4383
  indices depending on the substructure.\label{env}}
\end{figure}

In Fig. \ref{env} we present line strengths as a function of
environment. The latter is characterized in three different ways:
projected DM density (black circles), angular distribution in the
northern subcluster (blue squares), and angular distribution in the
southern subcluster (red squares) (see table \ref{tab_regs}). All
(passive and star-forming, but no AGN) galaxies have been stacked in
each region. We have separated galaxies in the northern subcluster
from those in the southern one aiming at taking into account the
merging nature of \CL\ \citep{drl05,gdr05}.

Clear trends of the age-sensitive indices with environment are
observed. No or little H$\delta_A$ in absorption ($<2$ \AA) is seen in
cluster areas with the $\Sigma_{DM} > 5 \times \sigma_{DM}$ (regions
17/HDMD and 18/MDMD), while values of H$\delta_A \sim 3$ \AA\ are
measured at projected densities $\Sigma_{DM} < 5 \times \sigma_{DM}$
(region 19/LDMD). The H6 index is observed to increase from $\sim3$
\AA\ to $\sim5$ \AA\ with decreasing projected DM density, while a
gradient in the opposite sense is observed for the D4000 index. In
terms of angular distribution, the H$\delta_A$ and H6 indices
increase, on average, toward the outskirts of each subcluster, while
the opposite trend is observed for the D4000 index. As opposed to the
EW(C4668), the EW(CN3883) and EW(Fe4383) show a notable decrease
toward lower DM density regions. Gradients with radial distance are
only notable for the CN3883 and Fe4383 indices depending on the
substructure.

\subsubsection{Morphology and stellar mass}

A behavior according to expectations is observed for the spectral
indices from different morphological stacks (see table
\ref{tab_ind1}). As we move from early-type to late-type galaxies, on
average, Balmer indices increase in intensity, the 4000 \AA-break
weakens, and indices such as CN3883 and Fe4383 tend to decrease in
strength, except for C4668. While differences in EW(C4668) between
ellipical, S0 and spiral galaxies do not exceed $\sim3$ \AA, irregular
galaxies show, on average, EW(C4668) values as large as $\sim23$
\AA. The H6 index is observed to correlate with H$\delta_A$ and
anti-correlate with D4000. On the other hand, the [O$\mathrm{II}$]
index presents a marked increase when moving to late-type
morphologies, reaching EW([O$\mathrm{II}$])$\sim {}-31$ \AA\ when
co-adding spectra of irregular galaxies (see table \ref{tab_ind1}).

Table \ref{tab_ind1} shows index values for regions 14/RSHM through
16/RSLM formed by passive, RS galaxies grouped according to stellar
mass. The average stellar mass derived from the spectrophotometric
fitting to the stacked data (see \ref{sed_fit}) for each one of these
three regions is given in table \ref{tab_sfh}. These values are
consistent, within the errors, with those of
$(1.70\pm0.14)\times10^{11}$, $(0.50\pm0.04)\times10^{11}$ and
$(0.20\pm0.01)\times10^{11} M_{\odot}$ for regions 14/RSHM, 15/RSMM
and 16/RSLM, respectively, obtained from directly averaging the
stellar masses of the individual galaxies within each of those bins.

Balmer indices increase and the D4000 index decrease as we move from
the most massive bin (14/RSHM) to the less massive one
(16/RSLM). Qualitatively, the relative variations between these
indices are preserved with respect to those in color-magnitude space
in the RS (see \ref{spind_rs}). This is in agreement with the
expectation by which the stellar mass is the main responsible of the
integrated stellar spectrum of galaxies and, therefore, their
color-magnitude properties.

No clear variations of the metal absorption features with stellar mass
is observed. Although the C4668 index seems to increase with stellar
mass, a more uniform set of values is measured for CN3883 and Fe4383
in the considered mass range. We note that these estimates are subject
to large uncertainties, and that the metallicity spread covers a
reduced range of $\Delta Z = 0.4 \ Z_{\odot}$ (see \ref{metal_range}
below).

\subsection{Considerations about H$\delta_A$ and H6}

With respect to the H$\delta_A$ index, it is important to take into
account the following caveat. As reported in \citet{drl05}, the
H$\delta$($\lambda$4101.7) line redshifted to the cluster redshift of
$z\sim0.84$ is close ($\sim50$ \AA) to the atmospheric A-band feature
at $\sim7600$ \AA.  Although a standard telluric correction was
applied to remove the A-band feature, this correction may have
introduced an uncertainty as large as $\sim20$\% in the EW of
H$\delta$ for some of the individual spectra \citep{drl05}. 

The effects of this correction may have propagated in some extend to
the final co-added spectra, for that the results reported in
\S\ref{res_indx} have to be considered with caution. However, since
our error bar estimates are obtained by taking into account the r.m.s
flux in the pseudo-continuum windows of the line, any additional noise
introduced by the telluric correction should thus be already included
in the error bars.

The observed variations of the H6 index with respect to H$\delta_A$
and D4000 suggest that the H6 line may be used as an indicator of
young stellar populations, whenever a significant young stellar
component is present. However, the use of this line to estimate ages
associated with a young ($\sim 1 - 2$ Gyr) stellar component is not
adviced unless a good modeling of metal lines and abudance ratios is
available. 

\begin{figure}
\plotone{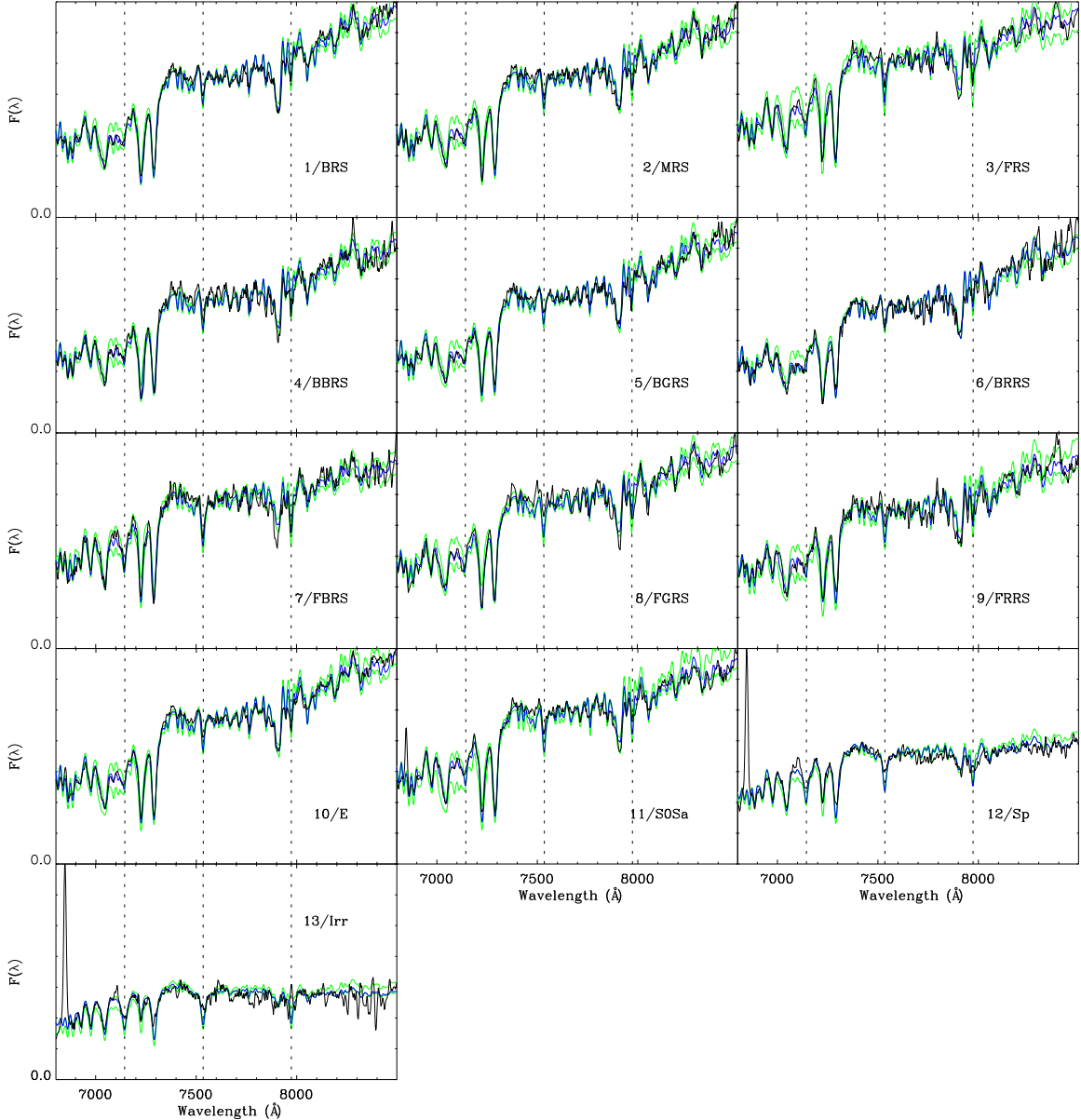}
\caption{The resulting stacked spectrum of galaxies for hyperspace
  regions 1/BRS through 13/Irr (as defined in table \ref{tab_regs}) is
  shown as a black line. The ``best-fitting'' spectrophotometric model
  (as defined in \S\ref{sed_fit}) is shown as the blue line, while the
  minimum and maximum fitting flux values within the intersection of
  the 99.7\% confidence regions (see \S\ref{sed_fit}), and as a
  function of wavelength, are represented by the green lines. Relative
  flux is in arbitrary units and all the spectra are shown redshifted
  to the mean cluster redshift, $z=0.837$ \citep{drl05}. In order to
  visualize the correlation between H6 and other Balmer features such
  as H$\delta$ and H$\gamma$, the vertical dashed lines indicate the
  location of these features from left to right,
  respectively.\label{specfits_a}}
\end{figure}

\begin{figure}
\plotone{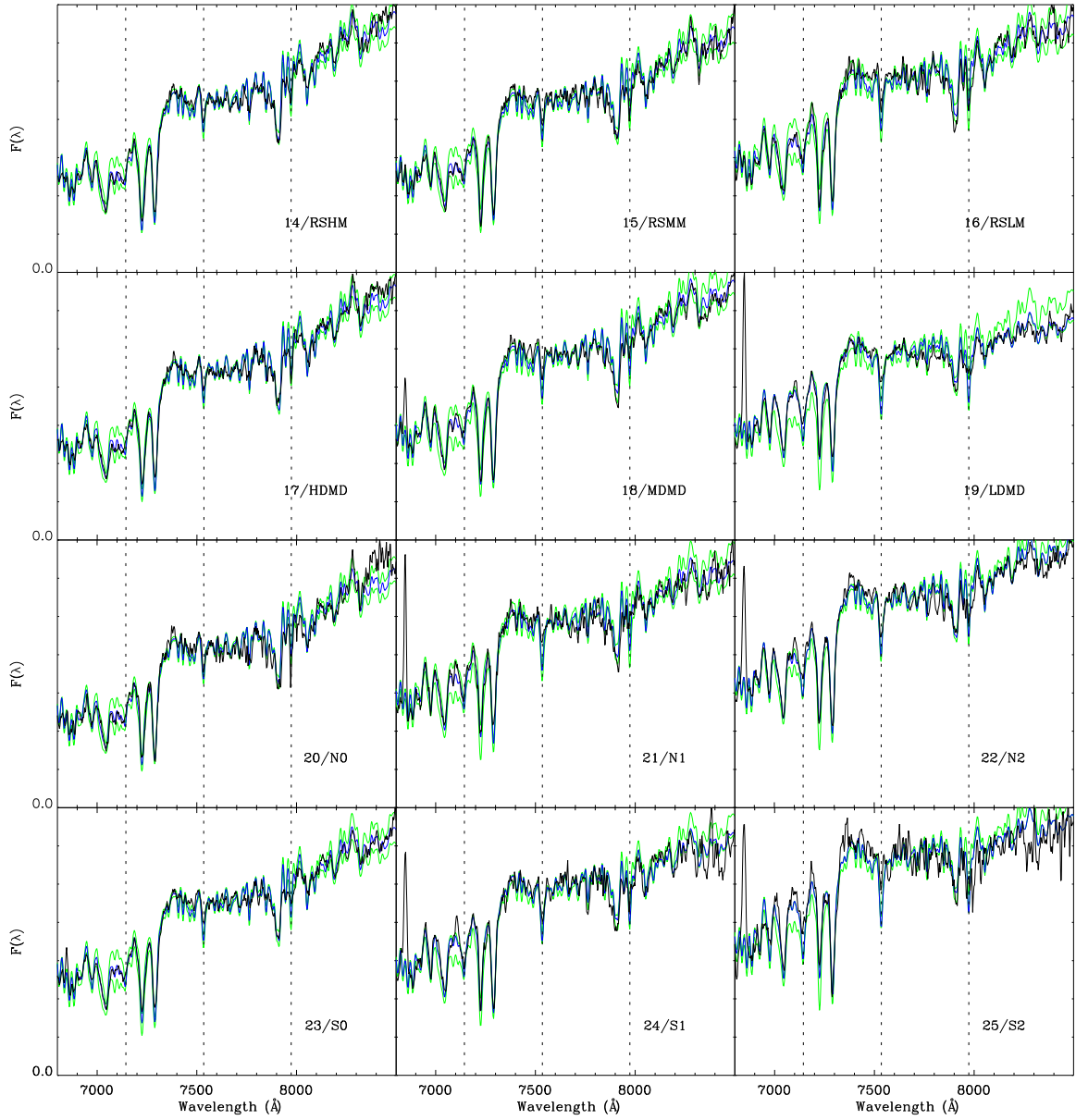}
\caption{As in Fig. \ref{specfits_a}, but here for hyperspace regions
  14/RSHM through 25/S2.\label{specfits_b}}
\end{figure}

In fact, our definition of the H6 index is based on a a single stellar
population, solar metallicity, 0.5 Gyr old model (see \ref{sp_indx})
which is not affected by strong metal features such as CN near 3883
\AA. The latter shows itself stronger in more massive, early-type
galaxies, thus affecting the pseudo-continuum and line windows of the
H6 index. For late-type galaxies, the CN feature decreases in
strength, and the H6 measurement becomes more representative of the
present younger stellar population. This can be seen in the different
panels of Fig. \ref{specfits_a} and Fig. \ref{specfits_b}. In order to
visualize the correlation between H6 and other Balmer features such as
H$\delta$ and H$\gamma$, the vertical dashed lines in
Fig. \ref{specfits_a} and Fig. \ref{specfits_b} indicate the location
of these features from left to right, respectively.

\subsection{Considerations about metallicity and internal dust}

\subsubsection{Metallicity range}\label{metal_range}

Local early-type galaxies are observed to have metallicities in the
range $\sim0.5-2 Z_{\odot}$ \citep*[see, e.g.,][]{k00,gcb06}, with
massive RS galaxies preferentially located at the super-solar end of
the metallicity distribution. Current observations suggest a little
evolution of the RS slope since $z\sim1.4$ down to $z=0$
\citep*[e.g.,][]{bfp03,bhf06,mhb06b,lrt08}, and the early
establishment of the bright end of the RS in gravitationally bound
systems \citep*[e.g.,][]{ktk07}. Hence, the metallicity of early-type
RS galaxies does not seem to have significantly changed between
$z\sim1$ and $z=0$, and the use of the local metallicity-stellar mass
relation in \citet{tmbm05} at $\sim0.84$, in order to infer metalicity
from stallar mass, thus seems to be justified.

Passive galaxies in the RS (bins 4/BBRS through 9/FRRS) have stellar
mass values of $8\times10^9 < M_{*,SED}<3\times10^{11} \ M_{\odot}$,
which, according to the local metallicity-stellar mass relation, gives
metallicity values of $1.2 < Z < 1.6 \ Z_{\odot}$. These metallicity
estimates are in agreement with the local values given in the previous
paragraph. We note that, considering the \citet{bc03} library, the
$1.2 < Z < 1.6 \ Z_{\odot}$ range is closer to solar metallicity than
to any other metallicity value available from that library using the
Padova 1994 tracks. This supports the original choice of using
\citet{bc03} solar metallicity models to calculate SFHs.

\subsubsection{Metallicity effects}\label{metal_effects}

It is important to make an assessment of the impact that metallicity
may have on our analyses and results. As metallicity is a function of
stellar mass \citep*[e.g.,][]{thk04,gcb06}, keeping the metallicity
fixed when fitting models to the observed data can induce systematic
errors in the stellar population parameters. This is especially
relevant when two galaxy samples are being compared, as a metallicity
difference between the galaxy populations can produce an apparent (and
spurious) difference in other properties of the stellar population,
such as age.

To quantify this bias, we carried out a test by which we fit a
\citet{bc03} model at solar metallicity to a series of subsolar and
supersolar metallicity \citet{bc03} models, assuming photometric and
spectroscopic errors consistent with the observed data.  The non-solar
metallicity models, spanning the range $0.5 - 2 \ Z_{\odot}$, were
obtained by interpolating a set of three models with 0.4, 1 and 2.5
$Z_{\odot}$. Unsurprisingly, the bias is significant for old models,
which have prominent metal features, while it is negligible for young
ones with spectra dominated by Balmer absorption. 

As in \citet{g09}, we define $\Delta T_{SFR}$ and $\Delta t_{fin}$ as
the difference between the best fitting solar metallicity models to
the solar metallicity input, and that of the best fitting solar
metallicity models to the nonsolar metallicity ones, for the mean star
formation weighted age and final formation time, respectively. For
models with $T-−t_{fin} \gtrsim 1.5$ Gyr, we found that $\Delta
T_{SFR}$ varies as $(0.7\pm0.1) \times Z/Z_{\odot}$ and $\Delta
t_{fin}$ as $(0.85\pm0.1) \times Z/Z_{\odot}$. Interestingly, at
supersolar metallicities, $\Delta T_{SFR}$ and $\Delta t_{fin}$ reach
a maximum for models whose age is $\sim2$ Gyr, and decrease for older
models. As this effect disappears when performing the same test on
\citet{m05} models, it is likely not intrinsic to the fitting
procedure (i.e., caused by the boundaries of the parameter grid, for
example) but due to a particularity of the \citet{bc03} templates. The
age at which the maximum difference occurs suggests that this is an
effect of the particular treatment of post-main sequence stars in the
\citet{bc03} model.

\subsubsection{Internal dust}

Another issue that must be addressed is the possible presence of
dust-enshrouded, star-forming galaxies in our red sequence bins. This
is especially relevant to the analysis of the faint-blue RS region
(7/FBRS) as it could mean that its composite spectrum is not that of a
quiescent stellar population with a significant contribution from
relatively young stars, but rather that of galaxies continuouisly
forming stars with a significant amount of internal dust. We test this
using extinction values derived from independent estimates of the star
formation rate in \CL\ obtained from Spitzer observations
\citep{mrr07}. Although we do not consider extreme cases of very low
dust and a very low [O$\mathrm{II}$] emission, the following
discussion strongly suggests that RS galaxies are not affected by
significant dust reddening.

As an estimate of the SFR in dusty galaxies in \CL\ we adopt the value
of $22^{+40}_{-10} \ M_{\odot} yr^{-1}$ derived by \citet{mrr07} from
deep 24$\mu$m observations with MIPS \citep{rye04} on the Spitzer
space telescope. We use the \citet{k98}'s relation between the
[O$\mathrm{II}$] emission at 3727 \AA\ and the SFR given by

\begin{equation}
SFR = (1.4\pm0.4)\times 10^{-−41} L([OII]) \ ,
\end{equation}

\noindent
where the SFR is units of $M_{\odot} yr^{-1}$ and the [O$\mathrm{II}$]
luminosity, in units of $erg \ s^{-1}$. The latter is related to the
rest-frame B-band luminosity and the EW([O$\mathrm{II}$]) as

\begin{equation}
L([OII])\sim(1.4\pm0.3)\times10^{29}\frac{L_B}{L_{B,\odot}}EW([OII]) \ .
\end{equation}

Together with these relations, we assumed the extinction curve derived
by \citet{cab00} for star-forming galaxies to estimate the ranges of
SFR and $E(B-V)$ values needed to produce an observed
EW([O$\mathrm{II}$]) of $-5$ \AA\footnote{We take this value as a
  fiducial one, considering \S\ref{sp_indx}.} given the rest-frame
B-band luminosity of the RS galaxies. This latter value and its
uncertainty were obtained from the best-fit $\tau$-models to the SED
of individual galaxies.

At $z\sim0.84$, the rest-frame B-band falls between the $i_{775}$ and
$z_{850}$ filters. The rest-frame B-band luminosity is, therefore,
strongly constrained by the observed SED and only weakly
model-dependent. We found that the amount of dust needed to damp the
[O$\mathrm{II}$] emission down to EW([O$\mathrm{II}$]) $=-5$
\AA\ results in an extinction of at least $E(B-V)=0.6$.

We also extended the grid of \citet{bc03} solar metallicity models
down to ages of $0.1$ Myr and reddened these model spectra using the
\citet{cab00} prescription, assuming the above putative
$E(B-V)\gtrsim0.6$ value and the SFR values from \citet{mrr07}. We
found that only models younger than $7$ Myr could reproduce the
observed colors in the RS when reddened. We then compared this subset
of models with the spectra of galaxies in the RS bright-red and
faint-blue bins (6/BRRS and 7/FBRS) using the D4000, H$\delta_A$ and
H6 indices. For most of the galaxies in those bins, neither the
reddened Balmer nor the D4000 model indices can reproduce the data,
with the model 4000 \AA-break being shallower.

The spectral features and color of one (possibly two) of the galaxies
in 6/BRRS is consistent with a dust-enshrouded, star-forming galaxy,
while no such dusty, star-forming galaxies were found in region
7/FBRS. We conclude that dusty star-forming galaxies do not form a
significant fraction of the population in the RS bins, but rather that
the average galaxy population at the RS faint, blue-end is truly a
quiescent one with a significant component of relatively young stars
and very little dust, if any.

\section{Discussion}\label{discussion}

The results presented in the previous sections can be used to deepen
our understanding of the effects that the cluster environment has on
the physical properties of cluster galaxies. Of special interest is to
understand the role that the environment plays in the assembly of the
RS in clusters.

The most prominent differences in SFHs and spectrophotometric
properties are found between the two opposite ends of the RS: regions
6/BRRS and 7/FBRS. On average, galaxies in the blue faint-end of the
cluster RS (7/FBRS) are $\sim1.8$ times younger and have $5$ times
less mass in stars than the average galaxy at the red bright-end of
the RS (6/BRRS). In terms of morphology, all of the galaxies in bin
6/BRRS are classified as early-type galaxies (E+S0) while 67\% (6/9)
of the galaxies in bin 7/FBRS are classified as such. The remaining
33\% of these sources is composed of one Sa galaxy, one spiral galaxy
with a morphology later than Sa, and one irregular galaxy. Possibly
only one of the galaxies in these two RS regions is a merger.

\begin{figure}
\plottwo{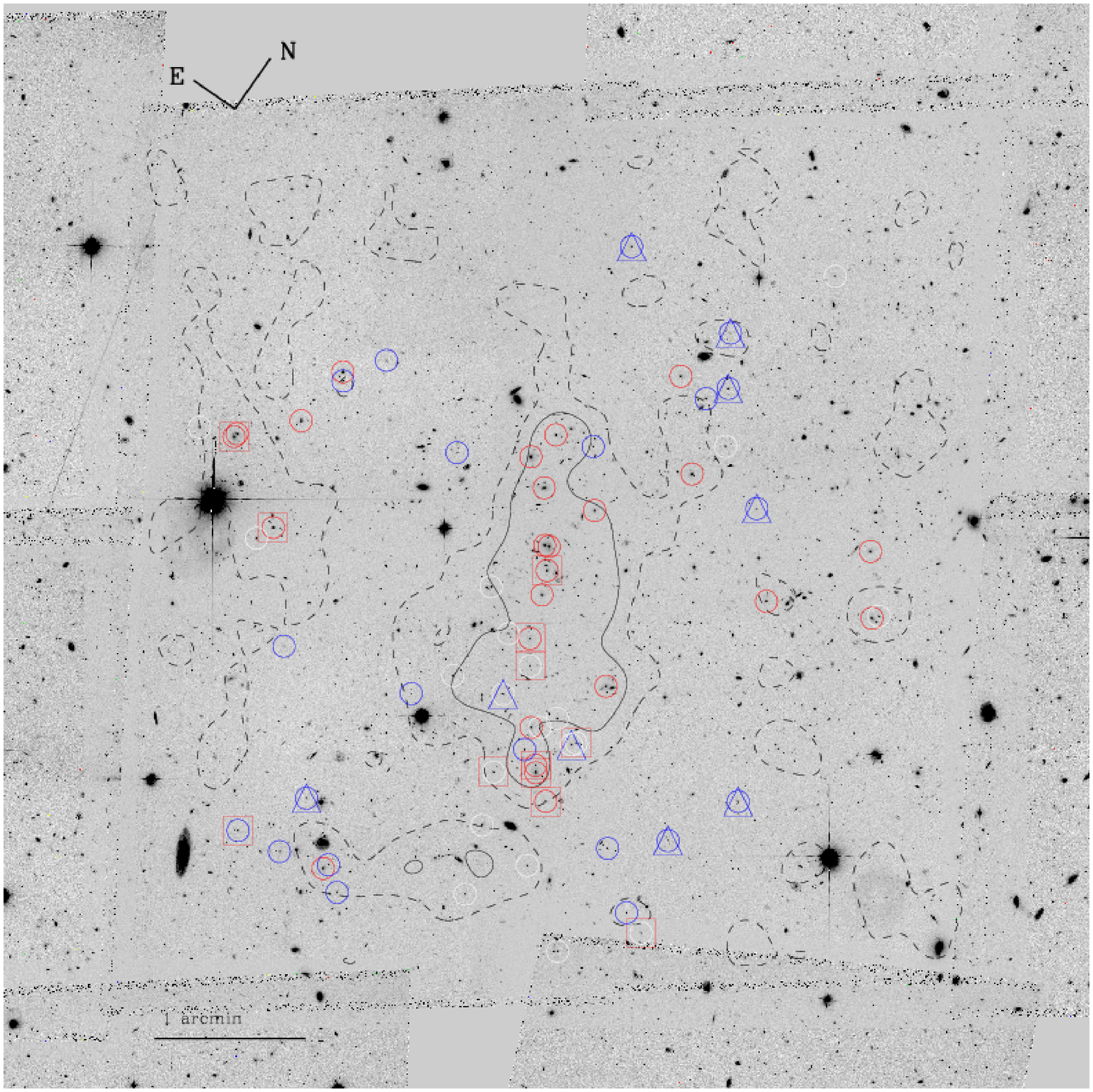}{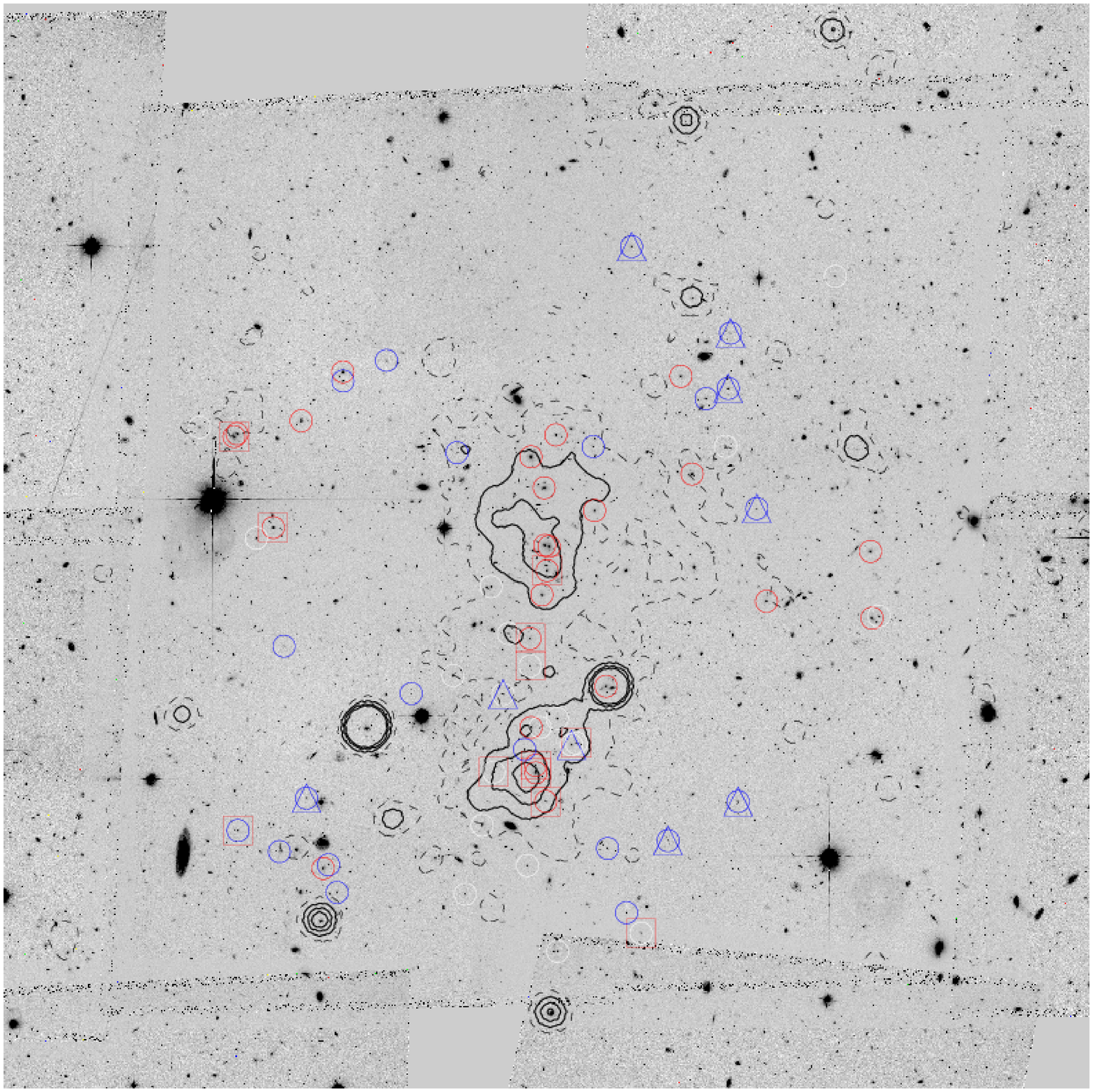}
\caption{Angular distribution of passive cluster members. The
  background image in each panel shows the same ACS data, 7\farcm2 a
  side and centered at the two brightest central galaxies of the
  northern clump, as in Fig. \ref{dmd_reg}. The {\it left} panel shows
  the same dark matter density contours of Fig. \ref{dmd_reg}. The
  {\it right} panel shows X-ray Chandra contours
  \citep*[see][]{drl05}. The dashed contour indicates the 3-$\sigma$
  level and the solid contours are the 10, 20 and 30-$\sigma$
  levels. The color symbols are the same for both panels. Red squares
  and blue triangles correspond to red-sequence passive members in
  regions 6/BRRS and 7/FBRS, respectively. The circles correspond to
  RS passive members grouped by stellar mass: red, white and blue
  symbols are galaxies within regions 14/RSHM, 15/RSMM and 16/RSLM,
  respectively, as defined in table \ref{tab_regs}. Clearly, less
  massive galaxies prefer lower density environment in the same way as
  the faintest and bluest galaxies in the red-sequence
  do.\label{dist8n9}}
\end{figure}

Fig. \ref{dist8n9} shows the location of RS passive members with
respect to both the projected DM density distribution (left panel) and
the gas distribution of the intracluster medium (ICM; right panel). In
both panels, circles correspond to cluster members grouped by stellar
mass: red, white and blue symbols are galaxies within regions 14/RSHM,
15/RSMM and 16/RSLM, respectively, as defined in table
\ref{tab_regs}. In addition, we identify the cluster members in
regions 6/BRRS and 7/FBRS as red squares and blue triangles,
respectively. The DM density contours are the same as those in
Fig. \ref{dmd_reg}, while the gas distribution is traced by the X-ray
Chandra contours \citep{drl05}.

It is clear, from Fig. \ref{dist8n9}, that a mass segregation exists
in the sense that the less massive galaxies prefer the lowest density
areas of the cluster, both in terms of projected DM density and gas
density. It is interesting to note that the southern subcluster is
populated by a significant number of galaxies in the intermediate mass
range (bin 15/RSMM). The most massive galaxies (those in region
14/RSHM) are preferentially located in the more massive northern
subcluster or near it, with a few of them populating the infalling
group to the East \citep{drl05,gdr05}.

More interestingly, there is a strong segregation in the location of
galaxies belonging to bins 6/BRRS and 7/FBRS. Most of the galaxies in
the bright, reddest bin (red squares) populate cluster areas where the
local DM density is $>5 \times \Sigma_{DM}$ and the X-ray emission is
detected at least at the 3-$\sigma$ level. Instead, most of the faint,
bluest passive galaxies in the RS (blue triangles) are situated in
cluster areas where the local projected DM density is below the $5
\times \Sigma_{DM}$ threshold and no significant X-ray emission is
detected.

The location of these passive, blue and faint RS galaxies in bin
7/FBRS within the cluster environment place interesting questions
about galaxy-environment interactions. In order to gain insight into
the environmental physics operating on these galaxies, we first
estimate the crossing time of the cluster. The time it would take to a
galaxy at a distance $r$ from the cluster center and moving at a speed
equal to the cluster's velocity dispersion, $\sigma_v$, to cross the
center of the cluster is given by $t_{cr} \simeq (r/$ Mpc$)
(\sigma_v/10^3$ km s$^{-1})^{-1}$ Gyr \citep{s88}. 

As seen in Fig. \ref{dist8n9}, most of the galaxies in 7/FBRS (blue
triangles) are located at $\sim1$ Mpc from the cluster center. At a
speed of $\sigma_v\sim1300$ km/s \citep{gdr05}, these galaxies would
have needed $\sim0.8$ Gyr to reach their observed position after a
first crossing through the center. Assuming that these galaxies in
region 7/FBRS originally come from a diametrically opposite point in
the cluster, ignoring projections effects, the total time spent inside
the cluster environment would thus be $\sim 2\times t_{cr}=1.6$ Gyr.

This time is shorter than $T_{SFR}$, but a factor of 2 larger than
the final formation lookback time, $T-t_{fin}$, estimated from the
average spectrum in bin 7/FBRS (see table \ref{tab_sfh}). We now
consider the scheme proposed by \citet{tek03} describing the action
range of a number of mechanisms likely responsible of altering galaxy
properties in the intracluster environment.

A first passage through the dens cluster core would have likely
suppressed star formation in a short timescale, resulting in a
lookback time $T-t_{fin}\lesssim1.6$ Gyr. Our estimate of
$T-t_{fin}\sim0.8$ Gyr (see table \ref{tab_sfh}) indicates that tidal
interactions with the cluster, ram pressure stripping and merging, all
effective at clustercentric distances $\lesssim1$ Mpc, may well have
been responsible of the spectrophotometric and morphological
properties of galaxies in bin 7/FBRS. In consequence, galaxies in
7/FBRS would be systems which are being accreted for the first time
into the cluster.

If galaxies in 7/FBRS are just entering the cluster environment for
the first time, according to \citet{tek03}, two other mechanisms, {\it
  harassment} \citep[e.g.,][]{mlk98} and/or {\it starvation}
\citep{ltc80,bcs02}, operating at large clustercentric distances
($\gtrsim1$ Mpc), may have contributed to shape the observed
properties of galaxies in bin 7/FBRS. Alternatively or in addition,
the quenching of star formation and the establishment of the
morphology and metal content may well have occurred within group
\citep[e.g.,][]{zm98,km08}, filament or neighboring ($>3$ Mpc) field
\citep{pkh08} environments outside the cluster. In fact, \CL\ is known
to be at the intersection of large scale filaments, in which
galaxy-galaxy interactions within groups may be the responsible
mechanism of the truncation of the star formation \citep{tka06}.

Furthermore, \citet{fww07} propose a ``mixed'' scenario to explain the
assembly and evolution of the RS, although their analyses include
galaxies in all environments, not only in clusters, and focus on the
bright-end of the RS, where their data are complete. In this scenario,
galaxies would enter the RS over a range of luminosities (masses),
first quenching their star formation via gas-rich (``wet'') mergers,
followed by some stellar (``dry'') mergers along the RS. In this way,
galaxies, once in the RS, would progressively move toward the bright,
red-end of it.

Our results on the SFH of RS galaxies (see table \ref{tab_sfh})
indicate that galaxies in the bright, reddest bin (6/BRRS) of the RS
of \CL\ quenched their star formation $\sim4$ Gyr prior to the epoch
of observation, while bluer galaxies of similar luminosity (bins
4/BBRS and 5/BGRS) continued forming stars down to a much recent
epoch. The reddest and brightest galaxies may have entered the RS when
being fainter (less massive), having $\sim4$ Gyr to reach their
current location in the RS. On the other hand, the amount of time
available for this to happen to bluer galaxies of comparable
luminosity is only $\sim2.5$ Gyr.

Studies of the redshift evolution of galaxy pair fractions and merger
rates \citep{lpk08} show that dry mergers, responsible of the creation
of massive, red galaxies, become as important as wet mergers only at
$z<0.2$ \citep[see also][]{v05}, being significantly surmounted by the
latter at $z\sim0.8$. An example of dry (red) mergers in a cluster
environment are those found in the cluster MS1054-03 at $z\sim0.83$
\citep{v99,tvf05}. Morphological signatures of dry merging are
expected to be visible for $\sim150$ Myr \citep{bnm06} which is
consistent with the lack of any interaction-driven feature in the
morphology of galaxies in 6/BRRS. Therefore, it is possible that
galaxies in the bright half ($18.5<K_s<20.75$) of the RS may have
entered it when being less luminous and reached their current location
after undergoing dry mergers.

However, no firm conclusion can be reached from the current data and
analysis. These massive, red galaxies may well have formed at
$z\gtrsim3$ through wet mergers in proto-cluster environments and
evolved without much interaction since. The amount and duration of any
wet merger for galaxies in these bright bins of the RS are limited,
however, to levels that are undetected by our data and analyses.

The situation of galaxies in the blue faint-end of the RS (7/FBRS) is
also of great interest. These galaxies seem to be transition objects,
entering the RS from the ``blue cloud''. They are passive and redder
than the ``blue cloud'', and would still contain young ($\lesssim 1$
Gyr) stars responsible of their bluer colors with respect to other RS
members such as those in bins 8/FGRS and 9/FRRS, which likely
entered the RS $\sim1$ Gyr earlier when they stopped forming
stars. The later evolution of galaxies in the faint half
($20.75<K_s<23.0$) of the RS, in particular those in 7/FBRS, may be
through dry mergers as suggested by \citet{fww07}, although at a small
rate ($\sim 10^{-4} h^3$ Mpc$^{-3}$ Gyr$^{-1}$) down to at least
$z\sim0.3$ \citep{lpk08}. Galaxies in 7/FBRS (and other bins) would
eventually move along the RS towards brighter and redder regions of
it, evolving, however, not necessarily in an entirely passive way as
suggested by \citet{jbd05}. This would imply some more extended star
formation episodes possibly triggered by mergers with some gas.

Eventually, the star-forming members in \CL\ that are closer to the
X-ray emitting gas \citep{drl05} and in the ``blue cloud'' will likely
suffer interactions with the denser ICM, leading to a suppression of
the star formation by ram pressure stripping. Alternatively, or in
addition to that, they will undergo wet mergers with other cluster
galaxies resulting in a rapid depletion of their gas content, which
would turn them into RS members.  At the same time, the ``blue cloud''
would also get replenished with infalling galaxies from the
surrounding field and from the large scale filaments connected to the
cluster \citep{tka06}. In the end, the result is that the RS becomes
extended toward fainter magnitudes as time goes by. The bright-end
builds up first, while fainter regions become progressively more
populated by galaxies migrating from the ``blue cloud'', such as those
in region 7/FBRS. This is consistent with a number of works until
present \citep*[e.g.,][]{dpa07,sse07}.

\section{Conclusions}\label{conclusion}

We have used a set of 134 galaxy spectra of cluster members and 5-band
photometry to investigate the stellar population properties of
galaxies in the $z=0.84$ cluster \CL. This very rich dataset allowed
us to study the variation of SFHs within the cluster as a function of
galaxy luminosity, color, morphology, photometric stellar mass, and
environment. These physical properties are used to define a hyperspace
of observables in which individual galaxies can be located. By
modeling both the co-added photometry and spectra inside specific
regions of this hyperspace, we can obtain useful information to
investigate the connection between environment and cluster galaxy
properties.

\CL\ is a dynamically young system \citep{gdr05}, consisting of two
massive substructures or subclusters in the process of merging, each
of them with a core dominated by massive, old early-type galaxies.  In
general, we find correlations of age ($T_{SFR}$) and final formation
lookback time ($T-t_{fin}$) with stellar mass (luminosity), color,
morphology, and environment, the latter characterized by local
projected DM density and projected location within the main
substructures (see table \ref{tab_sfh}).

For passive cluster galaxies in the RS, we find that they have formed
the bulk of their stars at $z_f\gtrsim2$ and stopped the star
formation at $z_{fin}\sim1$. We observe a range of these values, with
the red bright-end of the RS being characterized by $z_f\sim3.5$ and
$z_{fin}\sim2$, and the blue faint-end by $z_f\sim2$ and
$z_{fin}\sim1$. In terms of stellar mass, the most massive galaxies
($>8.4\times10^{10} M_{\odot}$) are observed to be $\sim1.5$ Gyr older
than the less massive ones ($<2.7\times10^{10} M_{\odot}$) under the
assumption of solar metallicity for both populations.

For the most massive (luminous) galaxies in the RS, we measure ages
$\gtrsim4.5$ Gyr with a short period of residual star formation
($\sim1.2$ Gyr), which suggest a formation scenario involving an
accelerated SFH and early quenching of star formation. This evolution
may have possibly been followed by further mass assembly via dry
mergers, and appears to be consistent with studies of higher redshift
clusters \citep*[e.g.,][]{bfp03,hse04,lrd04,mbs06a,tfk08}.

At the blue faint-end of the RS, however, we find a population of
galaxies that, while now passive, shows sign of having only recently
($T-t_{fin}\lesssim1$ Gyr) stopped forming stars. We also used our
stellar population modeling to discard the presence of dusty star
forming galaxies in our RS sample.

While stellar mass roughly follows the projected DM density, we find a
difference between the age-stellar mass and age-environment relations
which allowed us to partially distinguish the effects on the evolution
of cluster galaxies due to the ``intrinsic'' property of stellar mass
from those due to the cluster environment. Interestingly, we find that
the core of the southern subcluster is on average $\sim0.5$ Gyr
younger than that of the northern one, consistently with the former
being a less massive system \citep{drl05,gdr05,bhf06}.

In terms of age-sensitive spectral indices, qualitatively consistent
differences, with respect to those from the spectrophotometric
analysis, are observed between the average galaxy spectrum of the red
bright-end and that of the blue faint-end of the RS. The observed
variations of the H6 index with respect to H$\delta_A$ and D4000
suggest the H6 line as an indicator of young stellar populations,
whenever a significant young stellar component is present. However,
the use of this line to estimate ages associated with a young ($\sim1 -−
2$ Gyr) stellar component is not advised unless a good modeling of
metal lines and abundance ratios is available.

The general picture emerging from our spectrophotometric analysis is
one where low mass galaxies at the outskirts of the cluster may be
transformed from blue, star-forming to red, passive galaxies as they
fall into the cluster. The lack of a correlation between the
distribution of passive systems that recently stopped forming stars
and the x-ray emission may be interpreted in two ways: i) if these
galaxies are being observed after their first passage through the
cluster center, ICM-related interactions may be the responsible driver
for the quenching of the star formation; ii) if these galaxies are
first entering the cluster environment, the quenching of star
formation did not occur through interaction with the ICM but rather by
galaxy-galaxy interactions in the immediate vicinity of the cluster or
in nearby, infalling group, filament or field environments.

The ``downsizing'' of the star formation in cluster galaxies may thus
be an environmental effect, with the star formation time scale being
shorter than the assembly one (i.e., galaxies migrate from the ``blue
cloud'' to the the faint-end of the RS, with most of the mass assembly
happening later along the RS). On the other hand, the old age of the
brightest early-type galaxies are not incompatible with a ``monolithic
collapse'' scenario and it is likely that the quenching of star
formation happens through different mechanisms according to the epoch
and the environment \citep*[e.g.,][]{cnc07,fww07,rrn08}.

\acknowledgments

We thank the anonymous referee for useful comments and suggestions
that helped to improve the clarity and quality of this manuscript. We
also thank Harald Kuntschner and Marco Lombardi for valuable
discussions and suggestions. RD acknowledges the hospitality and
support of ESO in Garching, and the support provided by the BASAL
Center for Astrophysics and Associated Technologies and by FONDECYT
N. 1100540. CL acknowledges the financial support provided by the
Oskar Klein Center at the University of Stockholm. ACS was developed
under NASA contract NAS5-32865.

{\it Facilities:} \facility{HST (ACS)}, \facility{VLT (FORS)}, \facility{NTT (SOFI)}.


\clearpage
\begin{longtable}{lcccccc}
\tablecaption{Spectroscopically confirmed cluster members. Error in redshift is $\delta z\sim10^{-3}$.}
\tablewidth{0pt}
\tablehead{
\colhead{ID} & \colhead{R.A. (deg)} & \colhead{DEC. (deg)} & \colhead{$K_{s,Tot}$} & \colhead{$r_{625}-K_s$} & \colhead{z} & \colhead{E.L.}
}

18 & 28.1732 & -14.0068 & 20.359$\pm$0.700 & 3.146$\pm$0.023 & 0.8248 & 1 \\
26b & 28.2013 & -14.0062 & \nodata & \nodata & 0.8372 & 1 \\
35 & 28.1818 & -14.0057 & 21.707$\pm$0.104 & 2.980$\pm$0.265 & 0.8323 & 1 \\
47 & 28.1899 & -14.0038 & 20.471$\pm$0.067 & 3.230$\pm$0.066 & 0.8436 & 1 \\
67 & 28.1546 & -14.0020 & \nodata & \nodata & 0.8390 & 1 \\
81 & 28.1846 & -14.0017 & 20.738$\pm$0.068 & 3.432$\pm$0.063 & 0.8433 & 0 \\
85 & 28.1747 & -13.9980 & 19.515$\pm$0.039 & 3.404$\pm$0.050 & 0.8258 & 0 \\
97 & 28.1796 & -14.0001 & 21.213$\pm$0.086 & 3.103$\pm$0.088 & 0.8288 & 0 \\
113 & 28.1717 & -13.9988 & 21.477$\pm$0.096 & 3.152$\pm$0.088 & 0.8237 & 0 \\
125 & 28.2207 & -13.9976 & \nodata & \nodata & 0.8376 & 1 \\
129 & 28.1746 & -13.9973 & 21.580$\pm$0.105 & 3.098$\pm$0.106 & 0.8256 & 0 \\
131 & 28.2167 & -13.9712 & 19.197$\pm$0.035 & 3.698$\pm$0.044 & 0.8436 & 0 \\
144 & 28.1468 & -13.9963 & \nodata & \nodata & 0.8442 & 1 \\
145 & 28.1880 & -13.9978 & 23.333$\pm$0.233 & 1.573$\pm$0.287 & 0.8510 & 1 \\
161 & 28.1453 & -13.9960 & \nodata & \nodata & 0.8447 & 1 \\
177 & 28.1817 & -13.9938 & 21.441$\pm$0.100 & 2.766$\pm$0.107 & 0.8427 & 0 \\
182 & 28.1784 & -13.9940 & 21.524$\pm$0.099 & 3.337$\pm$0.118 & 0.8285 & 1 \\
184 & 28.1462 & -13.9947 & \nodata & \nodata & 0.8397 & 1 \\
204 & 28.1974 & -13.9904 & 19.022$\pm$0.032 & 3.833$\pm$0.045 & 0.8386 & 1 \\
234 & 28.1664 & -13.9880 & 21.143$\pm$0.087 & 2.458$\pm$0.097 & 0.8474 & 1 \\
241 & 28.1612 & -13.9889 & 20.970$\pm$0.078 & 3.340$\pm$0.079 & 0.8354 & 0 \\
248 & 28.2016 & -13.9891 & 20.630$\pm$0.066 & 3.267$\pm$0.069 & 0.8472 & 1 \\
258 & 28.2062 & -13.9752 & 19.658$\pm$0.042 & 3.772$\pm$0.045 & 0.8430 & 0 \\
267 & 28.1665 & -13.9884 & 20.824$\pm$0.074 & 3.345$\pm$0.091 & 0.8443 & 1 \\
270 & 28.1657 & -13.9872 & 20.027$\pm$0.053 & 4.007$\pm$0.071 & 0.8450 & 1 \\
291 & 28.1491 & -13.9862 & 21.081$\pm$0.081 & 3.347$\pm$0.080 & 0.8357 & 0 \\
295 & 28.1642 & -13.9844 & 20.576$\pm$0.066 & 2.282$\pm$0.079 & 0.8370 & 1 \\
300 & 28.1825 & -13.9836 & 19.962$\pm$0.048 & 0.598$\pm$0.053 & 0.8201 & 2 \\
306 & 28.2072 & -13.9744 & 20.004$\pm$0.050 & 3.497$\pm$0.058 & 0.8539 & 1 \\
327 & 28.1677 & -13.9828 & 22.499$\pm$0.151 & 0.484$\pm$0.153 & 0.8247 & 1 \\
328 & 28.1958 & -13.9837 & 22.100$\pm$0.132 & 2.822$\pm$0.142 & 0.8403 & 0 \\
332 & 28.1654 & -13.9822 & 20.405$\pm$0.059 & 3.155$\pm$0.063 & 0.8322 & 0 \\
344 & 28.1585 & -13.9818 & 20.829$\pm$0.071 & 3.189$\pm$0.071 & 0.8249 & 0 \\
347 & 28.1611 & -13.9818 & 21.801$\pm$0.113 & 2.365$\pm$0.098 & 0.8463 & 1 \\
377 & 28.1719 & -13.9786 & 20.569$\pm$0.064 & 3.045$\pm$0.073 & 0.8379 & 1 \\
387 & 28.1653 & -13.9739 & 19.636$\pm$0.044 & 3.716$\pm$0.048 & 0.8293 & 0 \\
394 & 28.1558 & -13.9758 & 21.253$\pm$0.089 & 2.818$\pm$0.098 & 0.8329 & 1 \\
396 & 28.1467 & -13.9778 & 22.436$\pm$0.154 & 2.952$\pm$0.137 & 0.8280 & 0 \\
397 & 28.1687 & -13.9771 & 20.599$\pm$0.064 & 3.347$\pm$0.062 & 0.8314 & 0 \\
418 & 28.1622 & -13.9754 & 20.396$\pm$0.059 & 3.587$\pm$0.060 & 0.8299 & 0 \\
432 & 28.2202 & -13.9732 & 20.453$\pm$0.061 & 3.327$\pm$0.070 & 0.8526 & 0 \\
436 & 28.1534 & -13.9742 & 21.092$\pm$0.082 & 3.106$\pm$0.080 & 0.8370 & 0 \\
439 & 28.1659 & -13.9733 & 19.556$\pm$0.049 & 3.604$\pm$0.070 & 0.8294 & 0 \\
445 & 28.1798 & -13.9728 & 20.225$\pm$0.054 & 3.221$\pm$0.057 & 0.8270 & 0 \\
455 & 28.1681 & -13.9730 & 21.413$\pm$0.094 & 3.360$\pm$0.082 & 0.8295 & 0 \\
468 & 28.1693 & -13.9707 & 19.757$\pm$0.044 & 3.492$\pm$0.053 & 0.8272 & 0 \\
474 & 28.1660 & -13.9715 & 21.676$\pm$0.105 & 3.019$\pm$0.104 & 0.8235 & 0 \\
477 & 28.1686 & -13.9700 & 21.069$\pm$0.081 & 3.370$\pm$0.077 & 0.8306 & 0 \\
491 & 28.1742 & -13.9703 & 20.974$\pm$0.076 & 2.989$\pm$0.068 & 0.8278 & 0 \\
498 & 28.1647 & -13.9689 & 20.774$\pm$0.071 & 3.041$\pm$0.086 & 0.8258 & 0 \\
509 & 28.1644 & -13.9684 & 20.713$\pm$0.070 & 3.470$\pm$0.075 & 0.8670 & 0 \\
511 & 28.1492 & -13.9689 & 21.782$\pm$0.111 & 2.835$\pm$0.092 & 0.8362 & 0 \\
513 & 28.1679 & -13.9680 & 20.504$\pm$0.062 & 3.284$\pm$0.062 & 0.8275 & 0 \\
522 & 28.2046 & -13.9674 & 21.545$\pm$0.105 & 1.536$\pm$0.117 & 0.8489 & 1 \\
543 & 28.2125 & -13.9647 & 19.267$\pm$0.035 & 3.389$\pm$0.044 & 0.8397 & 0 \\
547 & 28.1743 & -13.9660 & 20.377$\pm$0.058 & 3.626$\pm$0.060 & 0.8460 & 0 \\
548 & 28.1787 & -13.9652 & 20.292$\pm$0.056 & 3.354$\pm$0.058 & 0.8371 & 0 \\
551 & 28.1509 & -13.9634 & 19.711$\pm$0.043 & 3.525$\pm$0.050 & 0.8362 & 1 \\
557 & 28.1660 & -13.9613 & 18.555$\pm$0.025 & 3.246$\pm$0.026 & 0.8672 & 2 \\
571 & 28.1766 & -13.9639 & 20.051$\pm$0.051 & 3.494$\pm$0.060 & 0.8444 & 0 \\
595 & 28.1839 & -13.9629 & 20.880$\pm$0.073 & 3.122$\pm$0.074 & 0.8377 & 0 \\
598 & 28.1666 & -13.9616 & 19.004$\pm$0.032 & 3.594$\pm$0.044 & 0.8315 & 0 \\
626 & 28.1807 & -13.9288 & 20.897$\pm$0.073 & 2.889$\pm$0.071 & 0.8206 & 0 \\
648 & 28.1874 & -13.9432 & 21.435$\pm$0.095 & 3.022$\pm$0.098 & 0.8461 & 0 \\
650 & 28.1558 & -13.9431 & 20.524$\pm$0.063 & 2.197$\pm$0.077 & 0.8671 & 1 \\
654 & 28.1902 & -13.9442 & 19.738$\pm$0.043 & 3.634$\pm$0.050 & 0.8457 & 0 \\
663 & 28.1869 & -13.9439 & 21.583$\pm$0.101 & 3.117$\pm$0.087 & 0.8302 & 0 \\
679 & 28.1911 & -13.9496 & 18.613$\pm$0.026 & 3.595$\pm$0.035 & 0.8342 & 0 \\
682 & 28.1514 & -13.9477 & 22.848$\pm$0.172 & 2.907$\pm$0.150 & 0.8297 & 0 \\
688 & 28.1817 & -13.9488 & 19.812$\pm$0.045 & 3.419$\pm$0.056 & 0.8339 & 0 \\
701 & 28.1876 & -13.9509 & 18.817$\pm$0.029 & 3.506$\pm$0.036 & 0.8352 & 0 \\
735 & 28.1454 & -13.9527 & 22.811$\pm$0.177 & 2.574$\pm$0.164 & 0.8311 & 1 \\
805 & 28.1467 & -13.9604 & 21.783$\pm$0.115 & 2.423$\pm$0.116 & 0.8348 & 0 \\
811 & 28.2092 & -13.9610 & 21.306$\pm$0.089 & 2.820$\pm$0.089 & 0.8476 & 1 \\
851 & 28.1589 & -13.9270 & 19.861$\pm$0.046 & 2.923$\pm$0.055 & 0.8360 & 1 \\
859 & 28.1566 & -13.9084 & \nodata & \nodata & 0.8402 & 0 \\
868 & 28.1953 & -13.9088 & \nodata & \nodata & 0.8297 & 1 \\
889 & 28.1738 & -13.9090 & \nodata & \nodata & 0.8322 & 0 \\
895 & 28.1706 & -13.9097 & \nodata & \nodata & 0.8668 & 0 \\
898 & 28.1735 & -13.9094 & 14.416$\pm$0.001 & 8.727$\pm$0.016 & 0.8300 & 1 \\
928 & 28.1812 & -13.9117 & 20.836$\pm$0.072 & 3.264$\pm$0.070 & 0.8319 & 0 \\
931 & 28.2105 & -13.9533 & 21.139$\pm$0.082 & 3.086$\pm$0.088 & 0.8349 & 0 \\
1006 & 28.1742 & -13.9184 & 24.826$\pm$0.625 & 1.222$\pm$0.316 & 0.8485 & 1 \\
1067 & 28.1658 & -13.9223 & 22.360$\pm$0.145 & 2.956$\pm$0.136 & 0.8312 & 0 \\
1099 & 28.1999 & -13.9253 & 21.175$\pm$0.084 & 2.605$\pm$0.098 & 0.8230 & 0 \\
1102 & 28.1850 & -13.9243 & 22.094$\pm$0.128 & 2.743$\pm$0.121 & 0.8342 & 0 \\
1112 & 28.1486 & -13.9293 & 24.114$\pm$0.492 & 0.648$\pm$0.446 & 0.8666 & 1 \\
1131 & 28.1567 & -13.9279 & 24.444$\pm$0.423 & 0.945$\pm$0.290 & 0.8237 & 1 \\
1132 & 28.1714 & -13.9278 & 20.973$\pm$0.076 & 3.232$\pm$0.074 & 0.8259 & 0 \\
1146 & 28.1654 & -13.9300 & 22.606$\pm$0.169 & 1.583$\pm$0.165 & 0.8641 & 1 \\
1151 & 28.1817 & -13.9313 & 20.912$\pm$0.074 & 3.223$\pm$0.075 & 0.8330 & 0 \\
1172 & 28.1562 & -13.9304 & 19.571$\pm$0.040 & 3.143$\pm$0.046 & 0.8373 & 0 \\
1184 & 28.1855 & -13.9316 & 19.987$\pm$0.048 & 3.380$\pm$0.055 & 0.8288 & 0 \\
1204 & 28.1763 & -13.9335 & 20.739$\pm$0.068 & 3.234$\pm$0.074 & 0.8416 & 0 \\
1225 & 28.1820 & -13.9342 & 21.347$\pm$0.090 & 3.008$\pm$0.078 & 0.8399 & 0 \\
1226 & 28.1503 & -13.9348 & 20.323$\pm$0.058 & 3.109$\pm$0.062 & 0.8310 & 0 \\
1238b & 28.1576 & -13.9356 & 25.223$\pm$0.535 & 1.873$\pm$0.439 & 0.8456 & 1 \\
1239 & 28.1507 & -13.9355 & 19.184$\pm$0.034 & 3.179$\pm$0.041 & 0.8650 & 0 \\
1246 & 28.1688 & -13.9360 & 21.620$\pm$0.103 & 2.903$\pm$0.094 & 0.8306 & 0 \\
1258 & 28.1587 & -13.9410 & 19.162$\pm$0.034 & 3.381$\pm$0.044 & 0.8394 & 1 \\
1278 & 28.1768 & -13.9383 & 19.340$\pm$0.036 & 3.435$\pm$0.042 & 0.8215 & 0 \\
1290 & 28.2201 & -13.9398 & 20.390$\pm$0.060 & 2.217$\pm$0.070 & 0.8416 & 1 \\
1316 & 28.1654 & -13.9416 & 21.579$\pm$0.102 & 2.472$\pm$0.104 & 0.8456 & 1 \\
1338 & 28.1605 & -13.9425 & 20.074$\pm$0.053 & 3.307$\pm$0.063 & 0.8331 & 0 \\
1356 & 28.1888 & -13.9435 & 21.157$\pm$0.083 & 2.977$\pm$0.089 & 0.8290 & 0 \\
1367 & 28.1501 & -13.9422 & 20.288$\pm$0.056 & 3.265$\pm$0.064 & 0.8352 & 0 \\
1383 & 28.1951 & -13.9471 & 22.274$\pm$0.139 & 2.969$\pm$0.119 & 0.8334 & 1 \\
1386 & 28.1909 & -13.9459 & 20.179$\pm$0.053 & 3.399$\pm$0.060 & 0.8388 & 0 \\
1442 & 28.1791 & -13.9595 & 19.358$\pm$0.036 & 3.552$\pm$0.044 & 0.8318 & 0 \\
1454 & 28.1791 & -13.9572 & 21.707$\pm$0.111 & 2.720$\pm$0.118 & 0.8452 & 0 \\
1465 & 28.1807 & -13.9572 & 19.901$\pm$0.048 & 3.595$\pm$0.053 & 0.8365 & 0 \\
1466 & 28.1824 & -13.9552 & 19.119$\pm$0.035 & 3.448$\pm$0.045 & 0.8395 & 0 \\
1467 & 28.1829 & -13.9554 & 19.147$\pm$0.034 & 3.413$\pm$0.041 & 0.8412 & 0 \\
1483 & 28.1974 & -13.9550 & 21.174$\pm$0.084 & 3.254$\pm$0.084 & 0.8415 & 0 \\
1496 & 28.1439 & -13.9782 & 20.280$\pm$0.056 & 3.633$\pm$0.061 & 0.8300 & 0 \\
1499 & 28.2124 & -13.9583 & 21.229$\pm$0.086 & 3.123$\pm$0.075 & 0.8462 & 0 \\
1500 & 28.2131 & -13.9576 & 19.353$\pm$0.036 & 3.142$\pm$0.045 & 0.8477 & 0 \\
1501 & 28.2167 & -13.9707 & 19.244$\pm$0.034 & 3.408$\pm$0.044 & 0.8473 & 0 \\
1514 & 28.1352 & -14.0084 & \nodata & \nodata & 0.8264 & 0 \\
1530 & 28.2126 & -14.0052 & \nodata & \nodata & 0.8367 & 1 \\
1532 & 28.1837 & -14.0110 & 21.313$\pm$0.001 & 1.884$\pm$0.016 & 0.8413 & 1 \\
3013 & 28.1286 & -13.9623 & 12.764$\pm$0.001 & 12.314$\pm$0.051 & 0.8224 & 1 \\
3014 & 28.1263 & -13.9535 & 11.339$\pm$0.001 & 10.233$\pm$0.008 & 0.8474 & 1 \\
3015 & 28.2320 & -13.9473 & \nodata & \nodata & 0.8457 & 0 \\
5004 & 28.1303 & -13.9461 & 23.131$\pm$0.084 & 1.743$\pm$0.093 & 0.8292 & 1 \\
5010 & 28.1423 & -13.9383 & 22.933$\pm$0.209 & 2.011$\pm$0.221 & 0.8691 & 1 \\
5011 & 28.1434 & -13.9383 & 23.142$\pm$0.231 & 1.453$\pm$0.234 & 0.8691 & 1 \\
5015 & 28.2067 & -13.9774 & 21.081$\pm$0.080 & 3.189$\pm$0.083 & 0.8460 & 0 \\
5020 & 28.2252 & -13.9800 & 19.967$\pm$0.052 & 2.849$\pm$0.063 & 0.8387 & 1 \\
5027 & 28.1986 & -14.0130 & \nodata & \nodata & 0.8381 & 1 \\
5034 & 28.1817 & -13.9774 & 22.434$\pm$0.149 & 2.974$\pm$0.149 & 0.8304 & 0 \\
5042 & 28.1825 & -14.0094 & \nodata & \nodata & 0.8287 & 1 \\
5049 & 28.1794 & -13.9080 & \nodata & \nodata & 0.8447 & 1 \\
5063 & 28.1954 & -13.9024 & \nodata & \nodata & 0.8402 & 0 \\
 
\label{tab_membs}
\end{longtable}


\clearpage
\begin{deluxetable}{cll}
\tablecaption{Regions in color-magnitude space within the RS\tablenotemark{a}, morphology\tablenotemark{b}, stellar mass\tablenotemark{c}, projected dark matter density\tablenotemark{d} and location within the cluster used to group galaxies for stacking.}
\tablewidth{0pt}
\tablehead{
\colhead{Reg ID} & \colhead{Region Definition} & \colhead{Comments}
}
 
\startdata
1 & $18.5 < Ks < 20.2 \wedge 2.3 < r-Ks < 4.5$ & Bright-end red-sequence (BRS) \\
2 & $20.2 < Ks < 21.1 \wedge 2.3 < r-Ks < 4.5$ & ``Middle'' red-sequence (MRS) \\
3 & $21.1 < Ks < 23.0 \wedge 2.3 < r-Ks < 4.5$ & Faint-end red-sequence (FRS) \\
 & & \\
4 & $2.3 < r-Ks \leq -0.22Ks+7.65 \wedge 18.5 < Ks < 20.75$ & Bright, blue red-seq. (BBRS) \\
5 & $r-Ks > -0.22Ks+7.65 \wedge r-Ks < -0.22Ks+7.85 \wedge 18.5 < Ks < 20.75$ & Bright, green red-seq. (BGRS) \\
6 & $r-Ks \geq -0.22Ks+7.85 \wedge 18.5 < Ks < 20.75$ & Bright, red red-seq. (BRRS) \\
7 & $2.3 < r-Ks \leq -0.22Ks+7.65 \wedge 20.75 < Ks < 23.0$ & Faint, blue red-seq. (FBRS) \\
8 & $r-Ks > -0.22Ks+7.65 \wedge r-Ks < -0.22Ks+7.85 \wedge 20.75 < Ks < 23.0$ & Faint, green red-seq. (FGRS) \\
9 & $r-Ks \geq -0.22Ks+7.85 \wedge 20.75 < Ks < 23.0$ & Faint, red red-seq. (FRRS) \\
 & & \\
10 & $T < -2$ & Elliptical (E) \\
11 & $-2 \leq T \leq 1$ & Lenticular (S0/Sa) \\
12 & $1 < T \leq 6$ & Spiral (Sp) \\
13 & $6 < T$ & Irregular (Irr) \\
 & & \\
14 & $8.4 \times 10^{10} M_{\odot} < M_* \leq 3.9 \times 10^{11} M_{\odot} \wedge  2.3 < r-Ks < 4.5$ & Red-sequence high-mass (RSHM) \\
15 & $2.7 \times 10^{10} M_{\odot} < M_* \leq 8.4 \times 10^{10} M_{\odot} \wedge 2.3 < r-Ks < 4.5$ & Red-sequence medium-mass (RSMM) \\
16 & $4.8 \times 10^{9} M_{\odot} < M_* \leq 2.7 \times 10^{10} M_{\odot} \wedge 2.3 < r-Ks < 4.5$ & Red-sequence low-mass (RSLM) \\
 & & \\
17 & $\Sigma_{DM} \geq 20 \times \sigma_{DM}$\tablenotemark{e} & High local mass density (HDMD) \\
18 & $5 \times \sigma_{DM} < \Sigma_{DM} < 20 \times \sigma_{DM}$\tablenotemark{f} & Medium local mass density (MDMD) \\
19 & $\Sigma_{DM} < 5 \times \Sigma_{DM}$\tablenotemark{f} & Low local mass density (LDMD) \\
 & & \\
20 & $R \leq 1\farcm0$, North & Central sector, north (N0) \\
21 & $1\farcm0 < R \leq 2\farcm0$, North & First sector, north (N1) \\
22 & $2\farcm0 < R \leq 3\farcm0$, North & Second sector, north (N2) \\
 & & \\
23 & $R \leq 1\farcm0$, South & Central sector, south (S0) \\
24 & $1\farcm0 < R \leq 2\farcm0$, South & First sector, south (S1) \\
25 & $2\farcm0 < R \leq 3\farcm0$, South & Second sector, south (S2) \\
 & & \\
\enddata
 
\tablenotetext{a}{Regions ID 1 to 9}
 
\tablenotetext{b}{Regions ID 10 to 13. Morphological types from \cite{pfc05}}
 
\tablenotetext{c}{Regions ID 14 to 16. Stellar masses from SED fitting (see section \S\ref{sed_fit})}
 
\tablenotetext{d}{Regions ID 17 to 19. Local dark matter density map from \cite{jwb05}}
 
\tablenotetext{e}{$\sigma_{DM}=0.0057\times\Sigma_c$ \citep*[see][]{jwb05,bhf06}}
 
\label{tab_regs}
\end{deluxetable}


\clearpage
\begin{deluxetable}{lcccl}
\tablecaption{Definitions of spectral indices.}
\tablewidth{0pt}
\tablehead{
\colhead{Index} & \colhead{Line window (\AA)} & \colhead{Blue-continuum window (\AA)} & \colhead{Red-continuum window (\AA)} & \colhead{Reference}
}
 
\startdata
EW(H6) & 3868.00 - 3908.00 & 3854.00 - 3866.00 & 3910.00 - 3922.00 & This work \\
EW(H$\delta_A$) & 4083.50 - 4122.25 & 4041.60 - 4079.75 & 4128.50 - 4161.00 & \cite{wo97}\\
EW([O$\mathrm{II}$]) & 3716.30 - 3738.30 & 3696.30 - 3716.30 & 3738.30 - 3758.30 & \cite{tfi03}\\
D4000\tablenotemark{a} &  -  & 3850.00 - 3950.00 & 4000.00 - 4100.00 & \cite{bmy99}\\
EW(CN3883) & 3780.00 - 3900.00 & 3760.00 - 3780.00 & 3900.00 - 3915.00 & \cite{dc94}\\
EW(CN$_2$) & 4143.38 - 4178.38 & 4085.12 - 4097.62 & 4245.38 - 4285.38 & \cite{wfg94}\\
EW(Fe4383) & 4370.38 - 4421.62 & 4360.38 - 4371.62 & 4444.12 - 4456.62 & \cite{wfg94}\\
EW(C4668)\tablenotemark{b} & 4635.25 - 4721.50 & 4612.75 - 4631.50 & 4744.00 - 4757.75 & \cite{wfg94}\\
\enddata
\tablenotetext{a}{Also referred to as D$_n$(4000) \citep*[see][]{khw03}}
\tablenotetext{b}{C4668=Fe4668 \citep*[see][]{wtf95}}
\label{tab_ind_def}
\end{deluxetable}


\clearpage
\begin{deluxetable}{ccccccccc}
\tabletypesize{\scriptsize}
\tablecaption{Star formation history (SFH) parameters and stellar masses from best-fit model spectrum according to regions in the RS\tablenotemark{a}, morphology\tablenotemark{b}, stellar mass\tablenotemark{c}, local dark matter density\tablenotemark{d}, and location within the cluster. See Table \ref{tab_regs} for the definitions of the regions used for co-adding spectra. Columns, from left to right, are mean star-formation-weighted age, formation redshift, final formation lookback time from $z=0.837$, and final formation redshift.}
\tablewidth{0pt}
\tablehead{
\colhead{Reg ID} & \colhead{$N_{stack}$\tablenotemark{e}} & \colhead{S/N\tablenotemark{f}} & \colhead{$T_{SFR}$ (Gyr)} & \colhead{$z_{f}$} & \colhead{$T-t_{fin}$ (Gyr)} & \colhead{$z_{fin}$} & \colhead{$M_*$ ($\times 10^{11} M_{\odot}$)} & \colhead{Comment}
}
 
\startdata
1 & 24 & 53.7 & $4.6^{+0.9}_{-0.9}$ & $3.4^{+10}_{-1.0}$ & $3.3^{+1.4}_{-1.1}$ & $1.9^{+10}_{-0.5}$ & $1.9^{+2.2}_{-1.4}$ & BRS\\
2 & 25 & 41.6 & $3.8^{+0.8}_{-0.9}$ & $2.5^{+1.0}_{-0.6}$ & $2.2^{+1.1}_{-0.8}$ & $1.4^{+0.7}_{-0.3}$ & $0.8^{+1.0}_{-0.6}$ & MRS\\
3 & 18 & 39.9 & $3.0^{+1.0}_{-1.0}$ & $1.9^{+0.9}_{-0.4}$ & $1.1^{+0.6}_{-0.4}$ & $1.1^{+0.2}_{-0.1}$ & $0.4^{+0.4}_{-0.2}$ & FRS\\
 & & & & & & & & \\
4 & 10 & 23.8 & $3.9^{+0.7}_{-0.8}$ & $2.6^{+0.9}_{-0.6}$ & $2.3^{+1.0}_{-0.8}$ & $1.5^{+0.7}_{-0.2}$ & $1.7^{+2.2}_{-1.4}$ & BBRS\\
5 & 15 & 52.3 & $4.5^{+0.9}_{-1.0}$ & $3.3^{+10}_{-1.0}$ & $3.2^{+1.5}_{-1.1}$ & $1.9^{+10}_{-0.5}$ & $1.6^{+1.7}_{-1.1}$ & BGRS\\
6 & 12 & 25.8 & $4.7^{+0.8}_{-0.9}$ & $3.6^{+10}_{-1.1}$ & $3.5^{+1.3}_{-1.1}$ & $2.1^{+10}_{-0.5}$ & $1.5^{+1.7}_{-1.2}$ & BRRS\\
7 & 9 & 37.5 & $2.9^{+1.0}_{-1.0}$ & $1.8^{+0.8}_{-0.4}$ & $0.7^{+0.5}_{-0.2}$ & $1.0^{+0.1}_{-0.0}$ & $0.3^{+0.4}_{-0.2}$ & FBRS\\
8 & 11 & 25.1 & $3.4^{+0.9}_{-0.9}$ & $2.1^{+1.0}_{-0.4}$ & $1.5^{+0.7}_{-0.5}$ & $1.2^{+0.3}_{-0.1}$ & $0.5^{+0.7}_{-0.3}$ & FGRS\\
9 & 10 & 30.8 & $3.9^{+0.9}_{-1.1}$ & $2.7^{+1.3}_{-0.8}$ & $2.4^{+1.3}_{-1.0}$ & $1.5^{+1.0}_{-0.3}$ & $0.5^{+0.6}_{-0.3}$ & FRRS\\
 & & & & & & & & \\
10 & 42 & 48.3 & $4.4^{+0.9}_{-1.0}$ & $3.1^{+5.5}_{-0.9}$ & $3.0^{+1.5}_{-1.1}$ & $1.8^{+2.7}_{-0.4}$ & $1.3^{+1.4}_{-0.8}$ & E\\
11 & 40 & 48.3 & $3.9^{+0.9}_{-1.0}$ & $2.7^{+1.3}_{-0.8}$ & $2.4^{+1.3}_{-1.0}$ & $1.5^{+1.0}_{-0.3}$ & $0.7^{+1.1}_{-0.6}$ & S0/Sa\\
12 & 18 & 27.4 & $3.2^{+0.6}_{-0.8}$ & $2.1^{+0.4}_{-0.4}$ & $0.3^{+0.1}_{-0.0}$ & $0.9^{+0.0}_{-0.0}$ & $0.4^{+0.5}_{-0.4}$ & Sp\\
13 & 8 & 19.1 & $2.7^{+0.6}_{-0.9}$ & $1.8^{+0.3}_{-0.4}$ & $0.1^{+0.0}_{-0.0}$ & $0.9^{+0.0}_{-0.0}$ & $0.2^{+0.3}_{-0.1}$ & Irr\\
 & & & & & & & & \\
14 & 25 & 52.9 & $4.6^{+0.9}_{-0.9}$ & $3.4^{+10}_{-1.0}$ & $3.2^{+1.4}_{-1.1}$ & $1.9^{+10}_{-0.5}$ & $1.9^{+2.1}_{-1.3}$ & RSHM\\
15 & 20 & 39.1 & $3.8^{+0.8}_{-0.9}$ & $2.5^{+1.0}_{-0.6}$ & $2.2^{+1.1}_{-0.8}$ & $1.4^{+0.7}_{-0.3}$ & $0.8^{+1.0}_{-0.6}$ & RSMM\\
16 & 21 & 62.0 & $3.1^{+1.0}_{-0.9}$ & $1.8^{+1.0}_{-0.4}$ & $1.2^{+0.6}_{-0.4}$ & $1.1^{+0.2}_{-0.1}$ & $0.4^{+0.5}_{-0.2}$ & RSLM\\
 & & & & & & & & \\
17 & 18 & 52.6 & $4.6^{+0.9}_{-0.9}$ & $3.4^{+10}_{-1.0}$ & $3.3^{+1.4}_{-1.1}$ & $2.0^{+10}_{-0.5}$ & $1.7^{+1.9}_{-1.2}$ & HDMD\\
18 & 33 & 37.3 & $3.8^{+0.8}_{-0.9}$ & $2.6^{+1.0}_{-0.7}$ & $2.2^{+1.1}_{-0.9}$ & $1.4^{+0.8}_{-0.3}$ & $0.8^{+1.1}_{-0.7}$ & MDMD\\
19 & 57 & 48.4 & $2.9^{+1.0}_{-1.0}$ & $1.8^{+0.8}_{-0.4}$ & $0.8^{+0.5}_{-0.3}$ & $1.0^{+0.2}_{-0.1}$ & $0.6^{+0.7}_{-0.4}$ & LDMD\\
 & & & & & & & & \\
20 & 14 & 26.1 & $4.5^{+0.9}_{-1.0}$ & $3.3^{+10}_{-1.0}$ & $3.2^{+1.5}_{-1.1}$ & $1.9^{+10}_{-0.4}$ & $1.5^{+1.7}_{-1.1}$ & N0\\
21 & 23 & 35.9 & $3.1^{+1.0}_{-0.9}$ & $1.9^{+0.9}_{-0.4}$ & $1.1^{+0.6}_{-0.4}$ & $1.1^{+0.2}_{-0.1}$ & $0.7^{+0.9}_{-0.5}$ & N1\\
22 & 18 & 42.8 & $2.9^{+1.0}_{-0.9}$ & $1.8^{+0.8}_{-0.3}$ & $1.0^{+0.6}_{-0.2}$ & $1.0^{+0.2}_{-0.0}$ & $0.8^{+1.1}_{-0.8}$ & N2\\
 & & & & & & & & \\
23 & 23 & 45.4 & $3.9^{+0.8}_{-0.9}$ & $2.7^{+1.1}_{-0.7}$ & $2.4^{+1.1}_{-0.9}$ & $1.5^{+0.8}_{-0.3}$ & $0.9^{+1.0}_{-0.6}$ & S0\\
24* & 14 & 31.1 & $3.1^{+1.0}_{-0.8}$ & $1.8^{+1.0}_{-0.3}$ & $1.2^{+0.6}_{-0.3}$ & $1.1^{+0.2}_{-0.1}$ & $0.5^{+0.7}_{-0.4}$ & S1\\
25 & 11 & 24.7 & $3.0^{+0.9}_{-1.0}$ & $1.9^{+0.7}_{-0.5}$ & $0.6^{+0.4}_{-0.1}$ & $1.0^{+0.1}_{-0.0}$ & $0.6^{+0.6}_{-0.5}$ & S2\\
 & & & & & & & & \\
\enddata
 
\tablecomments{Individual spectra have been weighted when co-added.}
 
\tablenotetext{a}{Regions ID 1 to 9}
 
\tablenotetext{b}{Regions ID 10 to 13. Morphological types from \cite{pfc05}}
 
\tablenotetext{c}{Regions ID 14 to 16. Stellar masses from SED fitting (see section \S\ref{sed_fit})}
 
\tablenotetext{d}{Regions ID 17 to 19. Local dark matter density map from \cite{jwb05}}
 
\tablenotetext{e}{Only spectra with a signal-to-noise greater than 3 have been co-added}
 
\tablenotetext{f}{Signal-to-noise measured within the wavelength interval defining the continuum windows for the H$\delta$ feature}
 
\tablenotetext{*}{No 3$\sigma$ intersection. Errors from stacked-spectrum only}
 
\label{tab_sfh}
\end{deluxetable}


\clearpage
\begin{deluxetable}{cccccccc}
\tabletypesize{\scriptsize}
\tablecaption{Spectral indices from co-added spectra according to regions in color-magnitude space\tablenotemark{a}, morphology\tablenotemark{b}, stellar mass\tablenotemark{c}, and local dark matter density\tablenotemark{d}, and location within the cluster\tablenotemark{e}. See Table \ref{tab_regs} for the definitions of the regions used for co-adding spectra.}
\tablewidth{0pt}
\tablehead{
\colhead{Reg ID} & 
\colhead{EW([OII])} & \colhead{D4000} & \colhead{EW(H6)} & \colhead{EW(H$\delta_A$)} & 
\colhead{EW(CN3883)} & \colhead{EW(Fe4383)} & \colhead{EW(C4668)}
}
 
\startdata
1 & $-0.27\pm0.55$ & $1.74\pm0.01$ & $2.00\pm0.96$ & $0.80\pm0.20$ & $10.76\pm2.08$ & $3.01\pm0.46$ & $5.56\pm0.83$ \\
2 & $2.61\pm0.55$ & $1.72\pm0.01$ & $3.59\pm0.58$ & $1.95\pm0.26$ & $15.55\pm1.66$ & $4.85\pm0.96$ & $4.43\pm1.06$ \\
3 & $-3.61\pm1.45$ & $1.52\pm0.02$ & $3.19\pm1.99$ & $-0.26\pm0.53$ & $24.22\pm4.25$ & $3.48\pm1.25$ & $8.08\pm1.48$ \\
 & & & & & & & \\
4 & $-0.07\pm0.56$ & $1.73\pm0.02$ & $3.14\pm0.77$ & $0.29\pm0.36$ & $19.40\pm1.27$ & $3.71\pm0.50$ & $5.33\pm0.94$ \\
5 & $-0.39\pm0.59$ & $1.71\pm0.01$ & $2.11\pm0.68$ & $0.75\pm0.20$ & $8.94\pm2.45$ & $3.06\pm0.70$ & $6.52\pm1.09$ \\
6 & $2.28\pm0.62$ & $1.82\pm0.02$ & $2.87\pm1.20$ & $1.59\pm0.39$ & $22.28\pm2.10$ & $4.75\pm1.12$ & $4.87\pm1.88$ \\
7 & $-2.16\pm0.77$ & $1.46\pm0.01$ & $5.87\pm0.49$ & $3.33\pm0.29$ & $7.96\pm2.98$ & $0.66\pm0.62$ & $2.98\pm1.78$ \\
8 & $-1.88\pm1.55$ & $1.66\pm0.04$ & $2.28\pm1.73$ & $-1.62\pm0.75$ & $26.55\pm3.49$ & $4.64\pm2.41$ & $9.23\pm1.99$ \\
9 & $1.44\pm0.61$ & $1.63\pm0.02$ & $3.19\pm1.08$ & $-0.34\pm0.35$ & $21.51\pm2.28$ & $5.52\pm0.88$ & $1.72\pm1.30$ \\
 & & & & & & & \\
10 & $1.07\pm0.61$ & $1.71\pm0.01$ & $3.06\pm0.64$ & $1.31\pm0.21$ & $14.68\pm1.61$ & $3.27\pm0.70$ & $2.46\pm0.72$ \\
11 & $-2.35\pm0.45$ & $1.61\pm0.01$ & $3.67\pm0.91$ & $0.56\pm0.22$ & $4.61\pm2.45$ & $1.43\pm0.85$ & $5.88\pm0.89$ \\
12 & $-22.63\pm0.63$ & $1.23\pm0.01$ & $4.25\pm0.55$ & $3.65\pm0.38$ & $-2.66\pm2.68$ & $0.10\pm0.44$ & $5.66\pm0.77$ \\
13 & $-30.56\pm0.46$ & $1.15\pm0.01$ & $6.50\pm0.37$ & $4.69\pm0.45$ & $4.26\pm2.00$ & $0.38\pm0.29$ & $22.63\pm5.77$ \\
 & & & & & & & \\
14 & $-0.27\pm0.54$ & $1.74\pm0.01$ & $2.03\pm0.96$ & $0.79\pm0.21$ & $10.75\pm2.09$ & $3.29\pm0.52$ & $6.37\pm0.83$ \\
15 & $2.83\pm0.44$ & $1.67\pm0.01$ & $3.89\pm0.46$ & $0.23\pm0.24$ & $15.21\pm1.30$ & $3.37\pm0.59$ & $7.98\pm0.81$ \\
16 & $1.60\pm0.72$ & $1.58\pm0.01$ & $5.03\pm0.75$ & $2.34\pm0.24$ & $12.64\pm2.31$ & $3.89\pm0.73$ & $1.15\pm0.81$ \\
 & & & & & & & \\
17 & $2.70\pm0.43$ & $1.74\pm0.01$ & $3.01\pm0.71$ & $-0.94\pm0.16$ & $19.48\pm1.58$ & $3.91\pm0.48$ & $1.44\pm0.75$ \\
18 & $-5.38\pm0.62$ & $1.62\pm0.01$ & $3.42\pm0.64$ & $1.24\pm0.28$ & $3.53\pm2.21$ & $3.28\pm0.74$ & $6.24\pm0.79$ \\
19 & $-10.86\pm0.44$ & $1.38\pm0.01$ & $5.00\pm0.62$ & $3.06\pm0.23$ & $-3.09\pm2.30$ & $0.67\pm0.40$ & $9.28\pm1.06$ \\
 & & & & & & & \\
20 & $0.06\pm0.51$ & $1.72\pm0.02$ & $2.99\pm0.62$ & $-1.08\pm0.33$ & $15.92\pm1.23$ & $2.32\pm0.63$ & $5.53\pm0.58$ \\
21 & $-11.47\pm0.45$ & $1.49\pm0.01$ & $4.82\pm0.73$ & $0.26\pm0.24$ & $12.58\pm2.12$ & $3.20\pm0.49$ & $4.74\pm0.93$ \\
22 & $-11.24\pm0.37$ & $1.46\pm0.01$ & $4.18\pm0.54$ & $3.92\pm0.20$ & $14.79\pm1.64$ & $1.13\pm0.53$ & $5.79\pm0.29$ \\
 & & & & & & & \\
23 & $0.61\pm0.42$ & $1.65\pm0.01$ & $3.78\pm0.63$ & $-0.11\pm0.19$ & $15.98\pm1.52$ & $2.37\pm0.47$ & $6.69\pm0.70$ \\
24 & $-12.98\pm0.56$ & $1.42\pm0.01$ & $4.86\pm0.58$ & $2.92\pm0.33$ & $-0.68\pm2.30$ & $3.48\pm0.68$ & $13.05\pm2.23$ \\
25 & $-9.23\pm0.65$ & $1.36\pm0.02$ & $6.04\pm0.58$ & $3.39\pm0.43$ & $-4.80\pm3.07$ & $0.38\pm0.72$ & $3.64\pm1.18$ \\
 & & & & & & & \\
\enddata
 
\tablecomments{Individual spectra have been weighted when co-added. Equivalent widths (EW) are in units of Angstr\"oms}
 
\tablenotetext{a}{Regions ID 1 to 9}
 
\tablenotetext{b}{Regions ID 10 to 13. Morphological types from \cite{pfc05}}
 
\tablenotetext{c}{Regions ID 14 to 16. Stellar masses from SED fitting (see section \S\ref{sed_fit})}
 
\tablenotetext{d}{Regions ID 17 to 19. Local dark matter density map from \cite{jwb05}}

\tablenotetext{e}{Regions ID 20 to 25}
 
\label{tab_ind1}
\end{deluxetable}

\clearpage


\end{document}